\documentclass[conference, 10pt]{IEEEtran}
\IEEEoverridecommandlockouts

\usepackage{cite}
\usepackage{amsmath,amssymb,amsfonts}
\usepackage{algorithmic}
\usepackage{graphicx}
\usepackage{textcomp}
\usepackage[table]{xcolor}
\usepackage{hyperref}

\def\BibTeX{{\rm B\kern-.05em{\sc i\kern-.025em b}\kern-.08em
    T\kern-.1667em\lower.7ex\hbox{E}\kern-.125emX}}

\usepackage{tikz}
\usepackage{glossaries-extra}
\usepackage{adjustbox}
\usepackage{siunitx}
\usepackage{subcaption}
\usepackage{siunitx}
\usepackage{cleveref}
\usetikzlibrary{shapes.geometric, math, arrows.meta,arrows, decorations.pathreplacing,positioning}
    
\setabbreviationstyle[acronym]{long-short}
\newcommand{\newac}[2]{\newacronym{#1}{#1}{#2}}

\newac{D2D}{Device-to-device}
\newac{M2M}{machine-to-machine}
\newac{P2P}{peer-to-peer}
\newac{QoS}{Quality-of-Service}
\newac{SIG}{Bluetooth Special Interest Group}
\newac{BLE}{Bluetooth Low Energy}
\newac{WFD}{Wi-Fi Direct}
\newac{RSSI}{Received Signal Strength Indicator}
\newac{API}{Application Programming Interface}
\newac{Tup}{D2D Uptime Communication}
\newac{NFC}{Near-Field Communication}

\begin{document}

\fbox{\begin{minipage}[b][1cm][c]{18cm}
\footnotesize This article has been accepted for publication in the Computer Communications Journal, Elsevier. This is the author's version which has not been fully edited and content may change prior to final publication. Citation information: \url{https://doi.org/10.1016/j.comcom.2022.07.019}
\end{minipage}}

\bstctlcite{IEEEexample:BSTcontrol}
\title{SLICES,  a scientific instrument for the networking community}

\author{\IEEEauthorblockN{Serge Fdida}
\IEEEauthorblockA{\textit{Sorbonne Université, CNRS, LIP6} \\
Paris, France \\
serge.fdida@sorbonne-universite.fr}
\and
\IEEEauthorblockN{Nikos Makris, Thanasis Korakis}
\IEEEauthorblockA{\textit{University of Thessaly} \\
Volos, Greece \\
\{nimakris, korakis\}@uth.gr}
\and
\IEEEauthorblockN{Raffaele Bruno, Andrea Passarella}
\IEEEauthorblockA{\textit{IIT, CNR} \\
Pisa, Italy \\
\{raffaele.bruno,a.passarella\}@iit.cnr.it}
\and
\IEEEauthorblockN{Panayiotis Andreou}
\IEEEauthorblockA{\textit{UCLan Cyprus} \\
Larnaka, Cyprus \\
pgandreou@uclan.ac.uk}
\and
\IEEEauthorblockN{Bartosz Belter}
\IEEEauthorblockA{\textit{Poznan Supercomputing and Networking Center}\\
Poznan, Poland \\
bartosz.belter@man.poznan.pl}
\and
\IEEEauthorblockN{Cedric Crettaz}
\IEEEauthorblockA{\textit{Mandat International}\\
Geneva, Switzerland \\
ccrettaz@mandint.org}
\and
\IEEEauthorblockN{Walid Dabbous}
\IEEEauthorblockA{\textit{Université Côte d'Azur, Inria}\\
Sophia Antipolis, France \\
walid.dabbous@inria.fr}
\and
\IEEEauthorblockN{}
\and
\IEEEauthorblockN{Yuri Demchenko}
\IEEEauthorblockA{\textit{University of Amsterdam}\\
Amsterdam, The Netherlands \\
y.demchenko@uva.nl}
\and
\IEEEauthorblockN{Raymond Knopp}
\IEEEauthorblockA{\textit{Communication Systems Dept., EURECOM}\\
Sophia Antipolis, France \\
raymond.knopp@eurecom.fr}
}

\maketitle

\begin{abstract}

A science is defined by a set of encyclopedic knowledge related to facts or phenomena following rules or evidenced by experimentally-driven observations. Computer Science and in particular computer networks is a relatively new scientific domain maturing over years and adopting the best practices inherited from more fundamental disciplines. The design of past, present and future networking components and architectures have been assisted, among other methods, by experimentally-driven research and in particular by the deployment of test platforms, usually named as testbeds. However, often experimentally-driven networking research used scattered methodologies, based on ad-hoc, small-sized testbeds, producing hardly repeatable results. We believe that computer networks needs to adopt a more structured methodology, supported by appropriate instruments, to produce credible experimental results supporting radical and incremental innovations. This paper reports lessons learned from the design and operation of test platforms for the scientific community dealing with digital infrastructures. We introduce the SLICES initiative as the outcome of several years of evolution of the concept of a networking test platform transformed into a scientific instrument. We address the challenges, requirements and opportunities that our community is facing to manage the full research-life cycle necessary to support a scientific methodology. 

\end{abstract}


\glsresetall

\section{Introduction}


Prototyping a scientific instrument to explore the design space of future and emerging digital infrastructures has often been considered as impossible or irrelevant. The reasons are manifold, i) it is hard to predict the future landscape and challenging scientific questions, ii) the technology is evolving too quickly, and iii) the community is fragmented.  Unfortunately, the recent COVID episode demonstrated that providing evidence is a difficult task, often grounded on experimental research, and that this process is long and complex but timely and absolutely necessary. 

Up to now, networking test platforms have tried to capture a variety of demands. Academia represents the first target group, necessitating  tools  to support the validity of the assumptions and provenance of results and data published in scientific papers.  However, very little has been done to cover the entire research and data lineage lifecycle to ensure the longevity of such data as well as its access to the wider research and innovation community. It is for that reason that the FAIR (Findable, Accessible, Interoperable, and Reusable) and Open Science Principles were developed promoting interoperability and reproducibility of the results. Industry forms a second target group that often emphasizes the value of solutions to support conformance or interoperability testing, or the importance to make test platforms available for SMEs or startup companies, because otherwise they will never have the opportunity to access such instruments. As a consequence, different target groups impose unique requirements and expectations that need to be addressed by current and future testbeds. 

The field has matured quite a lot over the last decades. The first phase of test platforms can be illustrated by facilities such as PlanetLab~\footnote{PlanetLab, https://planetlab.cs.princeton.edu/} and Orbit~\footnote{Open-Access Research Testbed for Next-Generation Wireless Networks (ORBIT), https://www.orbit-lab.org/}. In 2005, the concept of testbed federation was introduced and applied to PlanetLab with the deployment of PlanetLab Europe in 2007 \cite{planetlab}.  Since then, this concept has developed quite a lot~\cite{fed4fire},~\cite{federation}. Orbit's success~\cite{orbit} has been due to addressing the need for realistic environments to test wireless protocols that were becoming essentials. The second phase of test platforms was initiated in 2007 with the ambition of the NSF GENI \cite{geni} (\verb-120M$ 2008/2016-) and EU FIRE \cite{eu-fire} initiatives (\verb-200M€ 2007/2022-). This corresponds somehow to a more structured approach to build a research infrastructure for this domain. GENI’s approach was meant to design a nationwide test platform, composed of GENI Nodes and racks that the experimenters could program. In Europe, the ambition was to federate testbeds with very heterogeneous resources. Both initiatives were nicely articulated and produced SFA~\cite{sfa}, the Slice-based Federation Architecture, proposing a practical solution to federate the facilities managed  by independent authorities. The third phase already started with initiatives such as NSF PAWR in the US~\cite{pawr}, CENI in China and ICT 17/19/52 in Europe~\cite{5geve, 5genesis, 5gvinni}. Those are developed in parallel without much cooperation and collaboration at present. The novelty comes from two new types of stakeholders. The tech giants are developing their own facilities (experimental or production), which is providing a risk with respect to the competition with academic research as these private platforms are not open, neither are the data that they use to produce their results. Second, other initiatives have emerged supported by the open-source community, such as ONAP~\cite{onap}, ORAN~\cite{oran}, OpenAirInterface~\cite{oai} enabling new and unique opportunities to deploy fully programmable and virtualised network infrastructures.

As a scientific community, our first message is that we have to continue to raise global awareness and promote the importance of a scientific instrument, because it is a community responsibility. All the efforts highlighted above address the demand for the networking field, including the demand for the experimental validation of research results.
This experimental validation constitutes a cornerstone of any sound scientific methodology. Both for historical and practical reasons, though, experimentally-driven research in the networking field has been so far quite fragmented, characterised by the development of ``in-house" testbeds, with the purpose of validating specific innovations (of specific research groups). This has (i) limited the scale at which experiments can be executed, (ii) limited the reproducibility of results (another pillar of scientific research) due to customizations at the individual testbed level, and thus (iii) limited the credibility of results produced. Because digital infrastructures are rapidly becoming a fundamental technological basis of our society, we think this gap needs to be urgently filled, such that next generation networks (starting from beyond-5G and 6G) can be developed based on reference, large-scale experimental infrastructures that could act as reference point for the wide community of computer networks and distributed systems researchers.

In Europe, there exists a framework called ESFRI - European Strategy Forum on Research Infrastructures~\cite{esfri} that supports the design, implementation and operation of scientific instruments. ESFRI is fully driven by science and organized in phases that should guarantee the feasibility and sustainability of the instrument. This framework is used by all scientific domains with a similar vision about the objectives and methodology. Targeting a scientific instrument for our community means that we have to align and adopt the principles promoted by ESFRI. It starts with a clear statement about the scientific question that this particular instrument will address. For instance, does the Higgs boson exist? Formulating such a question, which is easily understandable by other disciplines, the stakeholders and citizens, is not straightforward in our domain.


In this paper, we present the SLICES ESFRI initiative that is meant to support the discovery process related to the future, emerging digital infrastructures. In~\Cref{section:The ESFRI framework}, we focus on the ESFRI framework defining the requirements to enter into the roadmap. It relates to the ability and value of the future facility as well its sustainability. \Cref{section:Design foundation principles} highlights some of the design issues that are still being debated. We illustrate these foundations with the example of a 5G network in~\Cref{section:5G}. From this analysis, we derive preliminary architecture guidelines for SLICES in~\Cref{section:Architecture}. The research lifecycle dimension is of utmost importance. We introduce EOSC as a valuable target to be articulated with SLICES and discuss the components of the full-research lifecycle in \Cref{section:research}. The interoperability with EOSC is presented in \Cref{section:Interoperability}. We illustrate this with an example borrowed from another discipline, presented in \Cref{section:Example}. Finally, we conclude in \Cref{section:conclusion} and list topics for future investigations.

\section{The ESFRI framework, a scientific instrument}
\label{section:The ESFRI framework}


ESFRI, the European Strategy Forum on Research Infrastructures, established in 2002, brings together national governments, the scientific community, and the European Commission, to support a coherent and strategy-led approach to policy making on Research Infrastructures (RIs) in Europe. The ESFRI Roadmap~\footnote{ESFRI Roadmap 2021, https://www.esfri.eu/esfri-roadmap-2021} contains the best European science facilities based on a thorough evaluation and selection procedure. The Strategy Report on Research Infrastructures 2021 includes the Roadmap 2021 and the ESFRI vision of the evolution of Research Infrastructures in Europe, addressing the mandates of the European Council, and identifying strategy goals. Since 2002, within the framework of ESFRI and the ESFRI Roadmap process, national governments have worked in close partnership with the European Commission and the scientific community to catalyse the establishment of over 50 European Research Infrastructures, mobilising investments of approximately €20 billion across the EU.

ESFRI applies a lifecycle approach to the development and implementation of RIs . The lifecycle concept describes the different milestones in the development, implementation and operation of a Research Infrastructure over time, specifying minimal key requirements that must be met to enter each stage. Application of this concept allows for a coherent assessment of the scientific and organisational maturity of Research Infrastructures across all fields of science.


The concept of a new RI typically emerges bottom-up from the scientific communities clustering around well identified scientific needs and goals. For RIs to remain relevant throughout the entire RI lifecycle, scientific excellence is the \emph{conditio sine qua non}, which becomes, together with adequate human resources, crucial when it comes to long-term persistence in the operational phase. Effective governance and sustainable long-term funding (public and private) are other key elements for ensuring long-term sustainability of RIs at every stage in their lifecycle. 

We observe that until 2018, ESFRI was organized in 5 WGs related to energy, environment, health and food, physical sciences and engineering, social and cultural innovation. We had to wait until 2018 to applaud the creation of a working group dealing with Data, Computing and Digital Research Infrastructures that clearly differentiates digital \emph{research} infrastructures (i.e., addressing the needs of the \emph{digital research} communities) from \emph{production} research e-infrastructures defined as ICT support to other sciences.

SLICES is a distributed research infrastructure. It is organized with a central node and a set of distributed nodes. The central node hosts the governance and resources needed to steer and run the facility. As a European facility, the distributed nodes are hosted by different member states, covering different types of needs. At the time of writing, we have 15 countries supporting the effort~\footnote{Slices, https://slices-ri.eu/community/}. Industry is supporting but not directly involved in contributing to a common good.

\section{Design foundation principles}
\label{section:Design foundation principles}

The future generation of digital infrastructures is designed to be programmable, extendable and scalable. Key enabling technologies are now available to empower this important transformation. An important concept is network disaggregation whereby networking software is separated from the switching and/or routing hardware and broken down into functional components that can be more efficiently operated. Software Defined Networking (SDN) assumes programmable network devices in which the forwarding plane is decoupled from the control plane. In addition, the control plane is logically centralized in a software-based controller (“network brain”), while the data plane is composed of network devices (“network arms”) that forward packets. Network Function Virtualization (NFV) will deliver the promises of a software framework to the network, creating the need for an efficient and effective orchestration of the network resources. 

The primary technologies and solutions to address the requirements for the SLICES facility can be classified as described in the next subsections.

\subsection{Software Defined Network and Network Function Virtualization}

Following the major evolution of telecommunications networks with the adoption of the internet technology and the emergence of cellular networks, we are now facing a paradigm shift in the way that digital infrastructures are designed and operated. Indeed, recent advances in networking such as SDN and NFV~\cite{sdn-nfv-review} are changing the way network operators deploy and manage Internet services. SDN and NFV, together or separately, bring to network operators new opportunities for reducing costs, enhancing network flexibility and scalability, and shortening the time-to-market of new applications and services. On the one hand, SDN introduces a logically centralized controller with a global view of the network state. On the other hand, NFV allows to fully decouple network functions from proprietary appliances and to run them as software applications on general–purpose machines. It is a scalable approach as it gives the operators the ability to scale their network architecture across multiple servers to adapt quickly to the changing needs of their customers. 

\subsection{Network Slicing}
Another disruptive concept that should help in realizing the vision is network slicing, which allows a single physical network to be segmented into multiple isolated logical networks of varying sizes and structures tailored to different types of services and customers~\cite{2018_cst_slicing}. It is a multi-tenant virtualization technique in which the various network functionalities are extracted from the hardware and/or software components and then offered in the form of slices to the different users of the infrastructure (tenants). Basically, each slice includes a number of dedicated physical resources and network functions, which are isolated from other slices and provide specific functionalities including RAN and core network. Network slicing aims to offer operators the possibility of creating, in real-time and on-demand, various levels of services for different enterprise verticals, enabling them to customize their operations. In particular, it allows service differentiation with different QoS levels, reliability and security. Network slicing requires a continuous reconciliation of customer-centric service level agreements (SLAs) with infrastructure-level network performance capabilities. However, one of the main issues to solve is how to meet the requirements of different network services programmed from a single physical infrastructure. Autonomic (AI-empowered) and self-optimised management is needed to dynamically create, scale down or up, and reconfigure according to application demands~\cite{2018_jsac_nfv,2022_6g_slicing}. It is important to remark that the slicing concept is one of the key innovations in 5G since Release 15, supported by NFV and SDN techniques. Gradually, this concept is evolving from supporting primarily core-network resource provisioning, to more advanced (and forward-looking) paradigms, whereby also edge and far-edge devices' resources can be virtualised and provided as dynamic, on-demand components of the network infrastructure~\cite{6953022,dressler22}. However, an end-to-end network slice composed of sub-slices that belong to different technological
domains (RAN, core, edge/cloud), requires hierarchical and distributed management solutions to cope with the heterogeneity of the orchestration systems of different technological domains~\cite{2018_access_slicing, 2022_access_zero-touch}. 

\subsection{Network disaggregation}
Legacy aggregated networking devices have been developed and commercialized by vendors for decades. The term aggregation refers here to the vertical integration of software and specialized hardware components, bundled into a proprietary networking device. Network device disaggregation is the ability to source switching hardware and network operating systems separately. The term white box switches refer to switches built on commodity hardware that run different possible Network Operating Systems (NOS). This approach is putting pressure on the legacy aggregated networking vendors, but requires talented developers to build and grow the solution. This concept has been extended to the radio access network: RAN disaggregation~\cite{bonati2020open,o-ran-disaggregation} was specified by 3GPP~\cite{functional-splits} and detailed by the Open Networking Foundation (ONF)~\cite{wypior2022open} as an important step allowing for dynamic creation and lifecycle management of use-case optimized network slices. The idea here is to split the RAN protocol stack so that the individual components can be developed independently by different vendors. This horizontal disaggregation also enables distributed deployment of RAN functions in the network.

\subsection{Distributed Platform}
All the aforementioned techniques SDN/NFV, network slicing and disaggregation can be combined in a distributed platform to test advanced networking scenarios in realistic large-scale environments. This could be done by leveraging virtualized computing and networking resources in a flexible way to provide support for solutions based on the use-case, geography and experimenter choice. In such a distributed platform, the functions of the RAN nodes (the base stations) may be deployed as a "Central Unit", centralizing the packet processing functions and executed as Virtual Network Functions (VNFs) on commodity hardware in edge cloud locations. One or more "distributed units" performing the baseband processing functions as VNFs on commodity hardware with possible hardware acceleration and several "radio units" running the radio functions with specialized hardware on antenna sites. In a more general setting, different functions can be deployed on different sites in the network in order to realize the required flexibility and assess performance of the different split options.

\subsection{Control and User-plane Separation}
Another vertical disaggregation consists in the separation of Control and User Planes (CUPS)~\cite{cups}. In fact, with the densification of the next generation radio access networks, and the availability of different spectrum bands, it is more and more difficult to optimally allocate radio resources, perform handovers, manage interfaces, and balance load between cells. It is therefore necessary to adopt centralized control of the access network in order to increase system performance. This approach can be realized by decoupling the intelligence from the underlying hardware in all parts of the network.


\subsection{Research Data Management}
SLICES wants to fully endorse and adopt the Open Science and FAIR principles, acting as a catalyst to enable and foster cutting edge research, data-driven science and scientific data-sharing. Several design considerations should be taken into consideration, including: 
(i) \emph{easy and open access} to scientific data to facilitate further knowledge discovery and research transparency ensuring the longevity of the data and access to the wider research and innovation community.
(ii) \emph{scalable architecture} to efficiently leverage a large number of storage resources to  support efficient data storage and compute, including highly parallelized data workflows to support experiments;
(iii) \emph{privacy preservation} methods for ensuring end-to-end security and privacy in compliance with relevant legal frameworks; and 
(iv) \emph{data quality assurance} methods to ensure data quality across multiple dimensions, such as accuracy, completeness and integrity, in order to improve data utility.
To address this, SLICES requires to carefully design and develop efficient and scalable data management, analysis and reporting mechanisms, supported by appropriate metadata profiles to cater for access and reuse of FAIR data and services. These tools need to capture and report the entire data lineage/provenance across the data management lifecycle, while also providing the systematic means for secure and trustworthy interoperability of data and services ensuring the authenticity and immutability of the shared data.

\section{Illustration with 5G}
\label{section:5G}


As an illustration, we are presenting how these concepts are transforming the design and operation of cellular networks. The evolution of 5G networks introduces architectural changes in the Radio Access Network (RAN) \cite{ghosh20195g} and the Core Network (CN) \cite{rommer20195g} that will have adverse impact on how research infrastructure testbeds are designed as to support a variety of use cases. 

\subsection{Disaggregation of 5G RAN}

The 5G NR \cite{5g-nr-3gpp} defines a fully distributed Radio Access Network (RAN), by breaking traditional radio components into Radio Units (RUs), Distributed Units (DUs), and Centralized Units (CU). On the contrary, 4G provides a small level of disaggregation between Remote Radio Units (RRUs) and the centralized Baseband Units (BBU).

5G RAN has evolved from 4G with significant improvements in capabilities and functionalities. With the usage of a wider range of carrier frequencies that includes part of millimeter wave (mmWave) frequency spectrum, and flexible frame structure with variable number of symbols per subframe, 5G NR can utilize up to 400MHz of bandwidth per carrier. Several platforms exist that implement the 5G stack fully in software. By making use of Software Defined Radios (SDR), such platforms can turn commodity equipment (e.g., with General Purpose Processors) to fully functional base stations. The two most prominent solutions in open source to implement such functionality are: 1) the OpenAirInterface5G platform (OAI) \cite{oai}, and 2) the srsRAN platform \cite{srs-ran}. Both platforms support the basic operations for the 5G NR, though OAI has a wider user base and implements more features, such as disaggregated operation for the RAN, several different supported SDRs, etc. From an architecture perspective, 3GPP Release 15 has introduced CU/DU split (3GPP Option 2 split \cite{functional-splits}) along with a Virtualized RAN architecture.  Splitting the higher layers of 3GPP software stack (SDAP, PDCP and RRC) and lower layers (RLC, MAC and PHY) into separate logical units, known as Centralized Unit (CU), Distributed Unit (DU) and Radio Unit (RU), enables to deploy them at separate locations. Further split of gNB-CU is induced by separation between the Control Plane (CP) and User Plane (UP) named as gNB-CU-CP and gNB-CU-UP. The NG-RAN Network Resource Manager (NRM) \cite{5g-nrm} was designed to enable “separate” provisioning of CU, DU, CU-CP, CU-UP. 

Possible options for decomposition of the RAN environment are studied, resulting in the identification of eight options (3GPP Options 1-8). Building on top of the different disaggregation options, and especially delving into the CP/UP separation (CUPS), Open RAN (O-RAN) architecture defines open and standardized interfaces among the different elements of the disaggregated RAN. Through the use of such standardized interfaces, interoperability of functions between different vendors is made possible, while programmability of the RAN through dedicated interfaces is enabled \cite{o-ran-controller}. O-RAN Alliance is responsible for an additional split of the CU-CP into Radio Intelligence Controller (RIC) and remaining part of CU-CP. O-RAN defines the specifications for interface definitions between CU, DU, RU and RAN intelligent controller (RIC) that can be deployed at the edge of the network. Depending on the operation of the RIC and the programmable functions in the gNB, the RIC can operate in real-time mode ($< 1ms$ latency for programming the different functions, e.g., for Radio Resource Management) or near-real-time/non-real time mode (e.g., for the application and integration of Machine Learning models to the operation of the RAN). The OAI community works closely with the O-RAN ecosystem to ensure interoperability of key interfaces for experimentation with such disaggregated RAN topologies. Moreover, OAI aims for interoperability with upcoming O-RAN compliant radio-units to allow experimental infrastructure initiatives such as SLICES to make use of industry-grade radio solutions.

\begin{figure}[h]
\centering
\includegraphics{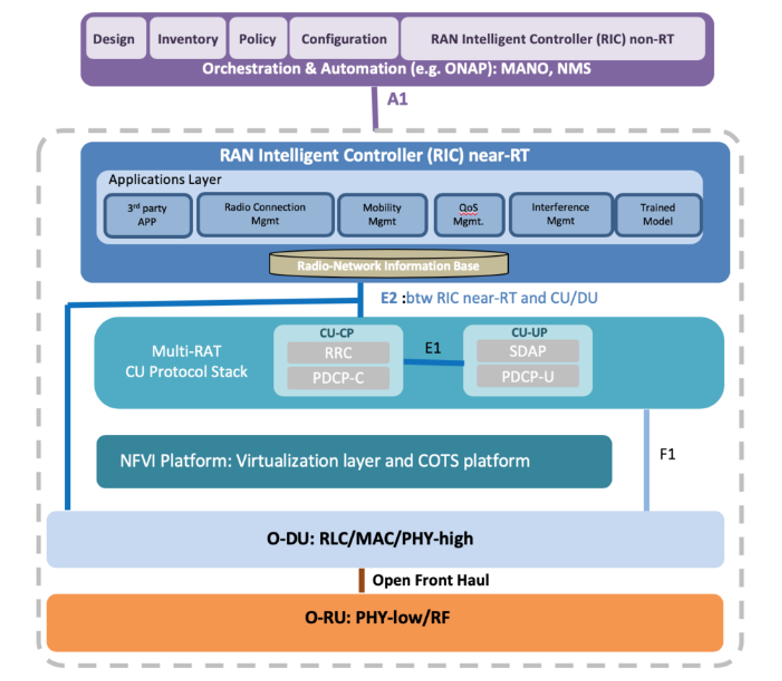}
\caption{Open-RAN deployment and programmable interfaces}
\label{fig:Open-RAN-deployment}
\end{figure}

It is reasonable to presume that the information model in O-RAN, as presented in \Cref{fig:Open-RAN-deployment}, will be the extension of the 3GPP NRM, with additional Managed Element object classes for RIC and possibly with extension of the information models for CU, CU-CP and DU. Similar to the O-RAN programmable interfaces, dedicated solutions for specific platforms exist, that open up the programmability of the RAN functions in practice. For example, the FlexRIC platform (also called as FlexRAN) \cite{flexric,flexran}, developed by Eurecom for OAI, allows the programmability of the OAI RAN in real-time, by exposing a REST interface. The interface can be used for retrieving statistics from the network as well, allowing for the advanced monitoring of the RAN in real-time. The FlexRAN controller is under further extension for becoming compatible with the O-RAN interfaces for programming the network.  Similar to the FlexRAN platform, the SD-RAN  platform developed by the Open Networking Foundation (ONF) is complementing O-RAN’s focus on architecture and interfaces by building and trialing O-RAN compliant open-source components. SD-RAN \cite{onf-sd-ran} is developing a near-real-time RIC (nRT-RIC) and a set of exemplar applications that run on top (xApps) for controlling the RAN. 
Towards integrating all the above efforts for the end-to-end deployment of the cellular network with extended use of virtualized services, the AETHER framework is currently under development by ONF \cite{onf-aether}. AETHER combines three main elements, namely, a control and orchestration interface to the RAN, an edge cloud platform (the AETHER edge), with support for cloud computing APIs, and a central cloud (the AETHER core), for orchestration and management. The AETHER project integrates several ONF efforts, including SD-RAN, ONOS\cite{berde2014onos}, CORD \cite{peterson2016central} and OMEC \cite{onf-omec}, for providing a fully-fledged solution for the deployment of the cellular network in an end-to-end manner. 

\Cref{tab:open-source-frameworks} lists out different open-source frameworks and projects that can be utilized to implement RAN and MEC infrastructure.

\begin{table}[h]
\centering
\begin{tabular}{|>{\columncolor{red!40!white!70}} p{0.5in} |p{0.4in}|p{1.1in}| p{0.9in}|}
    \hline
    \rowcolor{red!40!white!70}
      \textbf{Name}   &  
      \textbf{Network domain}  & 
      \textbf{Description}  & 
      \textbf{References/ links} \\\hline
      
      \textbf{OAI}\cite{oai}&
      RAN & 
      eNodeB, gNodeB and UE software &
         https://openair\newline interface.org\\\hline
      
      \textbf{srsLTE}\cite{srs-ran}&
      RAN &
      eNodeB, gNodeB and UE software &
      https://openair\newline interface.org/ \\\hline

      \textbf{SD-RAN}\cite{onf-sd-ran}  &
      RAN and Edge &
      Framework for RAN components and RAN intelligence controller &
      https://github.com/\newline srsran/srsRAN \\\hline
      
      \textbf{AETHER}\cite{onf-aether}&
      RAN and Edge &
      5G/LTE, Edge-Cloud-as-a-Service (ECaaS) &
      https://opennet\newline working.org/sd-ran/ \\\hline
      
      \textbf{FlexRIC}\cite{flexric}&
      RAN &
      Real-time controller for software-defined RAN &
      https://gitlab.\newline eurecom.fr/mosaic\newline5g/flexric \\\hline
\end{tabular}%

\caption{open-source frameworks and projects that can be utilized to implement RAN and MEC infrastructure}
\label{tab:open-source-frameworks}
\end{table}

\subsection{Disaggregation of 5G Core}

In 5GC, one of the most important characteristics is the separation of the User Plane (UP) functions from the Control Plane (CP) functions (3GPP TS 23.501 \cite{ts-23501}). UP functions mainly take care of traffic forwarding while the CP functions manage the authentication, network slice selections, etc. The principal advantage of such separation is being able to flexibly scale the CP functions independently on UP functions in case of traffic peak and vice versa. Another benefit lies in the flexibility to separately deploy CP functions so that some functions can be deployed, according to the requirement of the use case, in a centralized datacenter or a distributed one close to the RAN. The flexibility in scaling and deployment makes 5G networks more complex than previous generations of the telecommunication networks.

The Core Network (CN) is the central element of a network that provides services to customers who are connected to the access network. The 5G core network is referred as 5GC, and is an evolved version of EPC (LTE Evolved Packet Core network) as a cloud-native and service-based-architecture (SBA) \cite{arouk20205g}. The main components of the 5GC are the Access and Mobility Function (AMF), Session Management Function (SMF), User Plane Function (UPF), Unified Data Management (UDM), Authentication Server Function (AUSF), Policy Control Function (PCF), Network Exposure Function (NEF), Network Repository Function (NRF) and Network Slicing Selection Function (NSSF). These 5G network functions are cloud-native by design, thanks to the Service Based Architecture (SBA) design of the 5GC. Therefore, their instantiation can take place as Virtual Network Functions (VNFs) or Container Network Functions (CNFs) in any of the available virtualization platforms.


\begin{figure*}[t]
    \centering
    \includegraphics[width=0.8\textwidth]{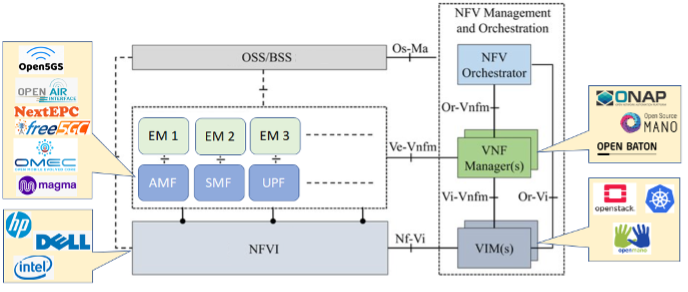}
	\caption{Cloud-native instantiation of the 5G Core Network}
	\label{fig:Cloud-native-instantiation}
\end{figure*}

The main goal is to adapt the 5GC functions independently when the load increases for any specific service or set of services, which is a major advancement from previous mobile network generations. To promote flexibility and reduce cost, it is possible to adopt COTS hardware at the NFV Infrastructure (NFVI) layer. These hardware resources are managed by open-source Virtual Infrastructure Management (VIM) software such as Openstack \cite{openstack}, OpenVIM \cite{openvim} or Kubernetes \cite{k8s}. The following \Cref{tab:open-source-solution} summarizes the available open-source solutions.

\begin{table}[h]
\centering
\begin{tabular}{|>{\columncolor{red!40!white!70}} p{0.8in} |p{0.4in}|p{0.9in}| p{0.8in}|}
    \hline
    \rowcolor{red!40!white!70}
      \textbf{Name}   &  
      \textbf{Network domain}  & 
      \textbf{Description}  & 
      \textbf{References/ links} \\\hline
      
      \textbf{Open5GS} \cite{open5gs}    &
      CN & 
      5G/LTE software &
         https://open5gs.org\\\hline
      
      \textbf{OpenAir\newline Interace CN (OAI-CN)} \cite{oai-cn}   &
      CN &
      5G/LTE software &
      https://openair\newline interface.org/ \\\hline

      \textbf{NextEPC}  \cite{nextepc}  &
      CN &
      LTE EPC software &
      https://nextepc.org \\\hline
      
      \textbf{srsEPC}  \cite{srs-ran}  &
      CN &
      LTE EPC software &
      https://github.com/\newline srsran/srsRAN \\\hline
      
      \textbf{Free5GC}  \cite{free5gc}  &
      CN &
      5G software &
      https://free5gc.org \\\hline
      
      \textbf{OMEC} \cite{onf-omec}   &
      CN &
      LTE EPC software &
      https://opennet\newline working.org/omec/ \\\hline
      
      \textbf{Magma} \cite{magma}   &
      CN &
      LTE/5G software &
      https://docs.magm\newline acore.org/docs/\newline basics/introduct\newline ion.html \\\hline
\end{tabular}
\caption{Available open-source solutions}
\label{tab:open-source-solution}
\end{table}

\subsection{Softwarization, orchestration, virtualization and programmability}

SDN is designed to make networks more flexible, controllable and agile. As a consequence, SDN enables network control to become directly programmable that makes its ability to provide network virtualization, automation, and create new services on top of virtualized resources. There exists a plethora of open source SDN solutions for mobile networks, including Open Networking Operating System (ONOS), Central Office Rearchitectured as a Datacenter (CORD), O-RAN, Open Network Automation Platform (ONAP) \cite{onap}, AETHER and SD-RAN.

\begin{figure}[!ht]
    \centering
    \includegraphics[width=\columnwidth]{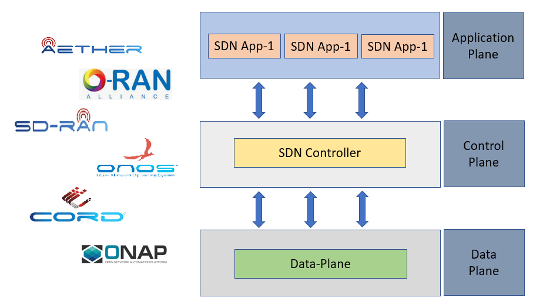}
	\caption{Key ONF SDN platforms}
	\label{fig:key-ONF-SDN}
\end{figure}

Management and Orchestration (MANO) frameworks \cite{ersue2013etsi} build on top of the network programmability and extended softwarization for network functions, and are being used to meet the agile and flexible management solutions for virtual network services in the 5G and beyond era. There are popular open source NFV MANO projects, namely OSM \cite{osm} and ONAP \cite{onap}.

\begin{figure}[!t]
    \centering
    \includegraphics[width=\columnwidth]{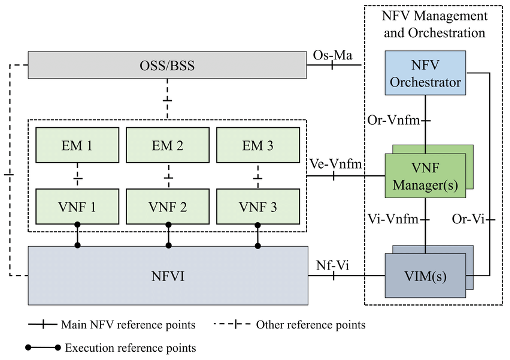}
	\caption{ETSI NFV-MANO architecture}
	\label{fig:ETSI ENF-MANO architecture}
\end{figure}

ETSI introduces the NFV MANO architecture, which comprises three main functional blocks, as further detailed below. MANO is an important component in managing the lifecycle of VNFs (including CFNs and PNFs) and hence managing overall infrastructure with agility and flexibility. The NFV MANO system entities, such as the Network Function Virtualization Orchestrator (NFVO), the Virtual Network Function Manager (VNFM) and the Virtual Infrastructure Manager (VIM), coordinate with each other over well-defined reference points to manage entities such as Network Functions Virtualization Infrastructure (NFVI), VNFs, CNFs, Physical Network Functions (PNFs) and Network Services (NSs). In the context of research testbeds, MANO framework provides efficiency by bringing network functions to several experimenters (tenants/users) at the same time. \Cref{fig:ETSI ENF-MANO architecture} illustrates these blocks with the reference points that connect them.

The three main components of the NFV-MANO architecture are detailed below:
\begin{enumerate}
    \item \textit{Virtual Infrastructure Manager (VIM)} performs controlling mechanisms for the NFV Infrastructure (NFVI) resources within an infrastructure provider. VIM is also responsible for receiving fault measurement and performance information of NFVI resources. Consequently, VIM can supervise NFVI resources allocation to the available VNFs;
    \item \textit{VNF Manager (VNFM)} conducts one or several VNFs and does the lifecycle management of VNFs. VNF lifecycle management involves establishing/configuring, preserving, and terminating VNFs;
    \item \textit{NFV Orchestrator (NFVO)} implements resource and service orchestration in the network. NFVO is split up into Resource Orchestrator (RO) and Network Service Orchestrator (NSO). First, the RO collects the current information regarding possible physical and virtual resources of NFVI through the VIM. Following this, the NSO applies a complete lifecycle management of multiple network services. In this way, the NFVO keeps updating the information about the available VNFs running on top of NFVI. As a result, the NFVO can initiate multiple network services. As part of the lifecycle management, the NFVO can also terminate a network service whenever no longer a service request is received for that specific service. In several solutions, NFVO and VNFM are integrated into the MANO section.
\end{enumerate}

Different frameworks have been developed in accordance with the NFV-MANO architecture, mainly aiming at providing fully-fledged solutions for the virtualized services lifecycle management. Such frameworks include multi-tenancy aspects, providing isolated slices of the infrastructure to each tenant, initially aiming at the execution of different vertical services on top of shared 5G infrastructure \cite{multi-tenant}. Such multi-tenancy aspects and isolation of traffic flows between each tenant of the infrastructure can be directly projected to the use of the same testbed infrastructure from multiple users concurrently, while providing guarantees for their performance. In  \Cref{tab:frameworks-for-VNF-lifecycle}, we list the different open source MANO frameworks that are currently widely utilized by the researchers as well as industry players like AT\&T, Telefonica and others. This table also  showcase the comparison between major open-source frameworks for VNF lifecycle management in terms of capabilities, multi-tenancy support, compliance or not with the NFV-MANO architecture, etc.

\begin{table}[h]
\centering
\begin{tabular}{|>{\columncolor{red!40!white!70}} p{1.3in} |p{0.3in}|p{0.3in}| p{0.4in}|p{0.5in}|}
    \hline
    \rowcolor{red!40!white!70}
      \textbf{Management and Orchestration framework}   & 
      \textbf{OSM} \cite{osm}& 
      \textbf{ONAP} \cite{onap}& 
      \textbf{CORD} \cite{peterson2016central}& 
      \textbf{OpenBaton} \cite{openbaton} \\\hline
      
      \textbf{Ease of Installation}   &
      X & 
      \checkmark &
      \checkmark &
      \checkmark \\\hline
      
      \textbf{Resource Footprint}   &
      High & 
      High &
      Medium &
      Medium \\\hline

      \textbf{Multi VIM support}   &
      \checkmark & 
      \checkmark &
      X &
      \checkmark \\\hline

      \textbf{VNF, CNF \& PNF Support}   &
      \checkmark & 
      \checkmark &
      \checkmark &
      \checkmark \\\hline
      
      \textbf{Multi-user Support (multi-tenancy)}   &
      \checkmark & 
      \checkmark &
      X &
      \checkmark \\\hline
      
      \textbf{Multi-site Support (multi- domain)}   &
      \checkmark & 
      \checkmark &
      \checkmark &
      X \\\hline
    
      \textbf{Network Slicing support}   &
      \checkmark & 
      \checkmark &
      \checkmark &
      \checkmark \\\hline
      
      \textbf{NFV-MANO compliance}   &
      \checkmark & 
      Partial &
      X &
      \checkmark \\\hline
     
\end{tabular}
\caption{Frameworks for VNF lifecycle management }
\label{tab:frameworks-for-VNF-lifecycle}
\end{table}


In the NFV world, containers are an emerging technology and the paradigm is standing between virtual machines and containers now. Containers show high utilization of computing resources and better performance than virtual machines. Multiple containers can be executed on the same host and share the same Operating System (OS) with other containers, each running isolated processes within its own secured space. Because containers share the base OS, the result is being able to run each container using significantly fewer resources than if each was a separate virtual machine (VM). Along with this trend, NFV industry has also been interested in the option of Containerized Network Functions (i.e., CNFs) instead of conventional Virtualized Network Functions (i.e., VNFs) due to its scalability and efficiency for operation and management. CNF-based solutions are also more appropriate for real-time networking functions. For those benefits, various mobile operators are trying to replace conventional VM-based NFV platforms with container-based platforms. 
Each VM includes a full copy of an operating system, the application, necessary binaries and libraries - taking up tens of GBs. VMs can be slow to boot, while Containers share the OS kernel with other containers, each running as isolated processes in user space. Containers take up less space than VMs (container images are typically tens of MBs in size), and thus handle more applications. Because they do not include the operating system, containers require fewer system resources and less overhead. They also tend to be faster to start/stop and they are ultra-portable across environments.


For low-latency use cases, 5G Core Network (CN) and RAN components are motivated to run as Containerized Network Functions (CNFs), instead of VMs in the case of Virtual Network Functions (VNFs), supported by tools like Kubernetes, that can deploy the services directly on bare-metal. Integration with the aforementioned NFVO tools like e.g., OSM is also possible. Open-source projects are moving towards cloud-native design, but until they become a reality, a mix of VNFs and CNFs could be adopted. Edge computing will have requirements for low-latency, cost-efficient infrastructure, secure with AI/ML capabilities. CNFs will be widely considered for the cases of Edge/Fog computing, due to the low complexity and fast instantiation of cloud-native services that can be achieved. However, simply forklifting existing 5G RAN software to a COTS platform is not enough. To realize the value of Cloud RAN, one needs to embrace cloud native architecture. Cloud native architecture facilitates RAN functions to be realized as microservices in containers over bare metal servers, supported by technologies such as Kubernetes. The \Cref{tab:open-source-container-solutions}  lists some of the widely used open-source container solutions.


\begin{table}[h]
\centering
\begin{tabular}{|>{\columncolor{red!40!white!70}} p{0.6in} |p{1.2in}|p{1.2in}|}
    \hline
    \rowcolor{red!40!white!70}
      \textbf{Container Solution}   &  
      \textbf{Description}  & 
      \textbf{References/ links} \\\hline
      
      \textbf{Kubernetes} \cite{k8s}  &
      Developed by Google, most widely used & 
      https://kubernetes.io/  \\\hline
      
      \textbf{Docker} \cite{docker}   &
      Software platform that allows you to build, test, and deploy applications quickly & 
      https://www.docker.com/   \\\hline
      
      \textbf{Openshift} \cite{openshift}    &
      Container management tool based on Kubernetes created by RedHat & 
      https://www.redhat.com/en/ technologies/cloud-computing/openshift    \\\hline
      
      \textbf{Apache Mesos} \cite{mesos}   &
      Apache Mesos is an open-source cluster management system & 
      http://mesos.apache.org/   \\\hline
      
\end{tabular}
\caption{Widely used open-source container solutions}
\label{tab:open-source-container-solutions}
\end{table}
For high-speed user-plane networking in the 5GC and advanced signal and information processing in the RAN, hardware accelerators are commonly used in industrial solutions to ensure real-time operation. For experimental network deployments, several open solutions in the context of AETHER and OAI can be now be leveraged to integrate P4 \cite{macdavidSOSR2021}, FPGA\cite{kaltenberger2021, oaixilinx2022} or GPU \cite{aerialSDK2022} based hardware accelerators in computing clusters. From a research perspective there are many challenges related to efficiently integrating such solutions with CNFs on generic computing platforms.

\subsection{Multi-access Edge Computing}

MEC (Multi-access Edge Computing)  has been developed as a solution for network operators, enabling the extension of telecommunications infrastructure with servers offering computing resources to users and service providers (including cloud service providers). The relevant standards are developed by the ETSI standardization organization within the ISG MEC (Industry Specification Group on Multi-access Edge Computing) working group. The ETSI MEC solution is now an integral part of the 5G network infrastructure, but it can also be used in LTE networks and other access networks. It should be emphasized that the MEC technique is a significant step in the development of telecommunications infrastructure towards a future, integrated communication and computing infrastructure.


Parallel to the standardization work, research is carried out on the specification of mechanisms and algorithms for MEC systems. In particular, as part of projects related to the implementation of NFV systems or orchestration in cloud systems, working groups were established to implement extensions of these systems and to support the computing technique at the network edge. Examples of such initiatives are: Open NFV Edge \cite{opennfv-edge}, Edge Automation through ONAP \cite{onap-project}, OpenStack Edge \cite{openstack-project}, or LinuxFoundation Edge \cite{lfedge-project}. The solutions proposed by the above projects are usually extensions of the architecture of orchestration systems developed for NFV backbone networks or cloud applications that offer the possibility of orchestrating applications at the edge of the network. Therefore, these solutions are not fully compatible with the ETSI MEC architecture. 

In addition, research projects aimed at developing prototypes of MEC systems fully compliant with ETSI standards have been developed \cite{5gcity-project},\cite{openness-project}. In particular, the SYMEC \cite{symec-project}, implemented an ETSI-compliant MEC platform on the low-energy ARM architecture.

\section{Architecture guidelines for SLICES}
\label{section:Architecture}

The initial guidelines on the various hardware building blocks for different types of SLICES facilities  can be categorized into four basic sub-systems:
\begin{itemize}
    \item Inter-Facility Interconnections and Intra-Facility Switching Fabric;
    \item Real-time and Non-real-time Computing;
    \item Radio Infrastructure;
    \item End-user devices. 

\end{itemize}
	
\begin{figure*}[t]
    \centering
    \includegraphics{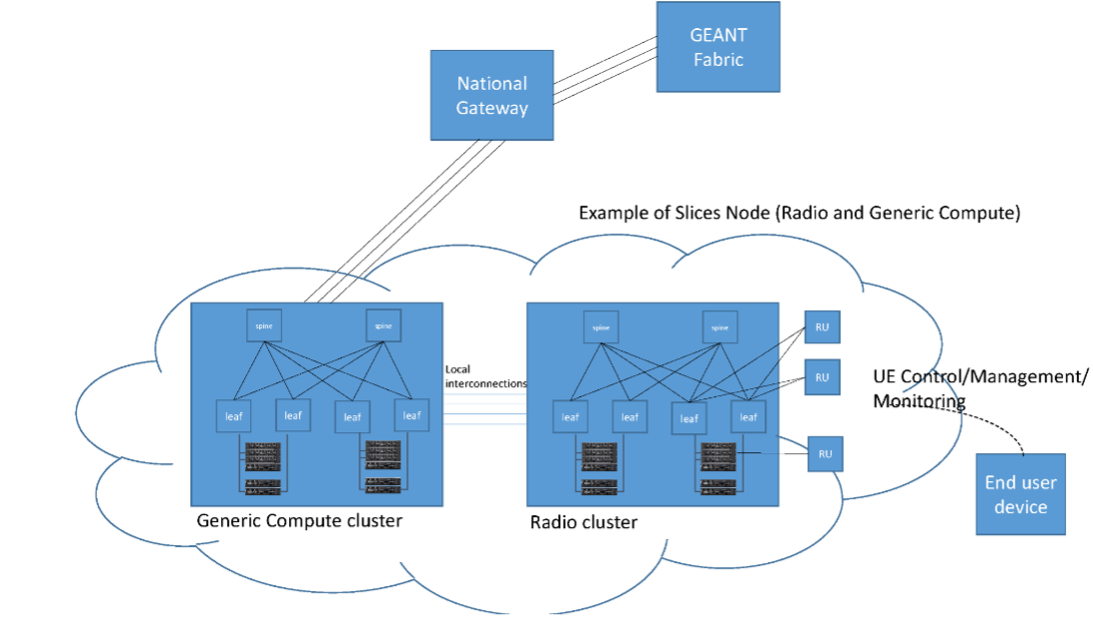}
	\caption{A high-level view of a SLICES node from an equipment standpoint}
	\label{fig:SLICES-node-view}
\end{figure*}

The example of a SLICES node is shown in \Cref{fig:SLICES-node-view}. It demonstrates two interconnected clusters in the same geographic region, one of which is equipped with Radio Units and the other is a more generic computing platform. 
The left cluster has a long-distance interconnection with the national gateway, which itself is interconnected with the GEANT fabric and the rest of the SLICES network. In the following subsections we provide some initial guidelines for the architecture of the various components.
As a general rule for hardware and network topologies, SLICES nodes should aim to mutualize as much as possible the types of computing and networking equipment in order to reuse deployment and configuration methods and to be able to share and establish common best practices. This follows the spirit of similar large-scale platform projects such as the Linux Networking Foundation OPNFV \cite{opnfv}, Cloud-Native Computing Foundation \cite{cncf} and the Open Compute Foundation \cite{ocf}.
Because of the lack of space and its diversity, we will not describe further the hardware components envisaged in SLICES.

\begin{figure*}[t]
    \centering
    \includegraphics[width=0.9\textwidth]{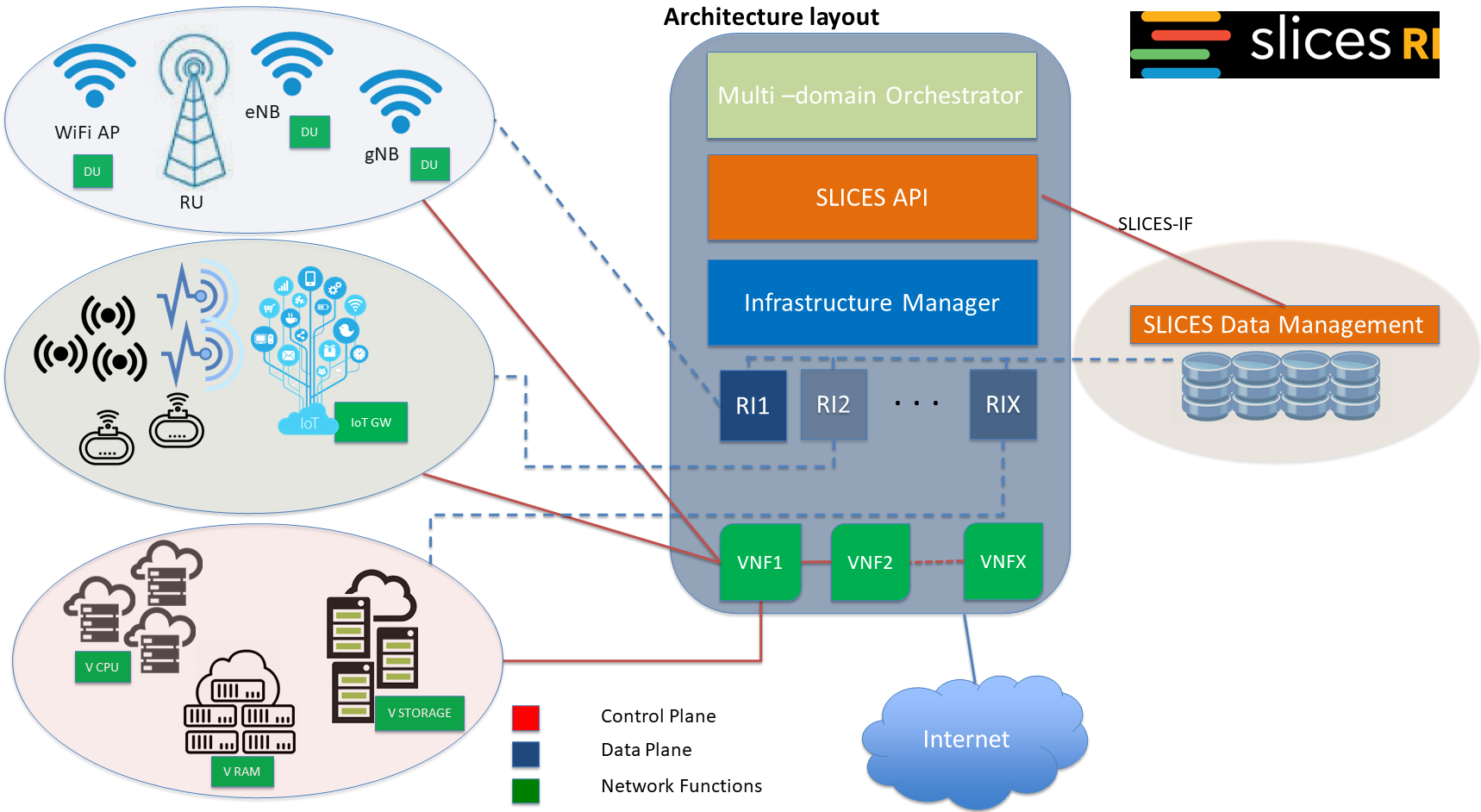}
	\caption{SLICES Infrastructure conceptual architecture}
	\label{fig:SLICES-conceptual-architecture}
\end{figure*}

As SLICES aspires to provide fully programmable remotely accessible infrastructure to the Digital Infrastructure community, the respective frameworks shall be developed for ensuring seamless and easy access to the experimental resources. The different site facilities will form an integrated single pan-European facility, adopting common tools for managing and orchestrating experiments over the infrastructure, as well as providing single access credentials to users. A first attempt to sketch our reference architecture, with respect to the tools used for its management, is described in \Cref{fig:SLICES-conceptual-architecture}.

Towards achieving this integration, the sites will adopt network virtualization for their disaggregated resources. Each node will be considered as a single domain for experimentation \cite{bernini2020multi}, while the overall orchestration of experiments will be performed through a centralized infrastructure. Site and node selection frameworks will be developed in the context of SLICES, towards ensuring the optimal use of resources among the sites. 

Moreover, and towards ensuring the smooth operation of the infrastructure, tools for facilitating access will be developed and deployed. Open-source software shall be employed, based on the paradigms of existing testbed access schemes, user authentication and authorization. This software will be appropriately tailored with new modules for managing the new equipment described in the previous section. 
\Cref{tab:frameworks-comparison} provides a comparison between the existing tools for the experimentation plane of the experiments, and the progress beyond them.

\begin{table*}[t]
\centering
\begin{tabular}{|>{\columncolor{red!40!white!70}} p{1.3in}|p{1in} |p{4.2in} |}
    \hline
    \rowcolor{red!40!white!70}
      \textbf{Existing Tools}   &  
      \textbf{Proposed Solution(s)}  & 
      \textbf{Benefits} \\\hline
      
      \textbf{Control and Management Framework (OMF)} \cite{omf}   &
      NFV-based orchestration solution & 
      Current tools provide metal as a service access to the testbed resources, or in some cases virtualized access by interacting with the respective VIM interface of a testbed. On top, the experiments can be orchestrated by using a publish/subscribe scheme for the communication between a centralized controller and the actual resources. Adopting an NFV-based solution will allow the orchestration of experiments as Virtual Network Functions (virtualized access) or Physical Network Functions (Metal as a Service access), through the adoption of industry-grade tools. These shall allow higher utilization of the testbed resources, increasing the user capacity of each testbed, more secure end-to-end experiments, end-to-end network configuration and experiment reliability.  \\\hline
      
      \textbf{SDN programmability}    &
      SDN Assist &  
      Current tools aim in providing a programmable interface for users that shall use their own controller for managing the flows in the network. In some cases, isolation of flows between different users on a switch is possible, through the adoption of tools like FlowVisor . Moving to an NFV-based orchestration solution supporting features like SDN Assist  enables the programming of flows for an experiment during the instantiation time. Based on an end-to-end programmable SDN plane (based on Open-vSwitch or hardware OpenFlow/P4 switches) programmability extends to the entire datapath used for the experiments, isolating users and providing multi-tenancy over the infrastructure.\\\hline
      
      \textbf{Wireless programmability}    &
      Open-RAN (ORAN) & 
      Current tools for programming the wireless components rely on specific interfaces dedicated to specific equipment for the RAN. As such interfaces become standardized, through efforts like O-RAN alliance, adopting such APIs can increase the supported equipment, and open-up more programmability for the RAN. As such tools use standardized interfaces, they integrate with several NFV based orchestration solutions, allowing a truly end-to-end experiment configuration and instantiation. \\\hline
      
      \textbf{Edge and Core Configuration}    &
      NFV-based orchestration solution & 
      Current tools provide Virtualized access to the core and cloud network configuration, or in some cases metal-as-a-service access. Switching to the same NFV based orchestration solution as the rest of the nodes will enable the seamless network configuration, and move the edge/core cloud configuration to supporting a different number of settings (such as cloud-native 5G network configuration).  \\\hline
      
\end{tabular}
\caption{Comparison of different proposed frameworks vs existing ones for the SLICES architecture}
\label{tab:frameworks-comparison}
\end{table*}


In terms of integration of the various components, the software tools shall encompass single-sign in procedures, with access certificates issued by a single authority. The resource discovery, reservation, and allocation shall comply with the access policies for SLICES (following the ESFRI principles) and be interchanged with the respective facility authorities through a standardized process. For this purpose, the SFA protocol \cite{sfa} has been extensively used in past and present solutions and  could inspire a future candidate together with new complementary or alternative proposals that will be considered as well. 


Based on the automation tools 
for accessing the infrastructure, 
we intend to equip new experimenters with a store in order to easily deploy services with a single click manner over the infrastructure. This can be achieved with these frameworks by using pre-compiled versions of services, and by supporting different methods for virtualization of resources (e.g., Virtual Machines, docker containers, Linux Containers). For example, public docker repositories provide different images that can be used to deploy commonly used services (e.g., databases, web services, applications and application servers) through a friendly interface. Moreover, the entire architecture will be augmented with the appropriate tools for experiment monitoring, experiment data and results visualization and cross-correlation analysis and inference with previous experiments executed over the infrastructure.

\begin{figure*}[!th]
    \centering
    \includegraphics[width=0.6\textwidth]{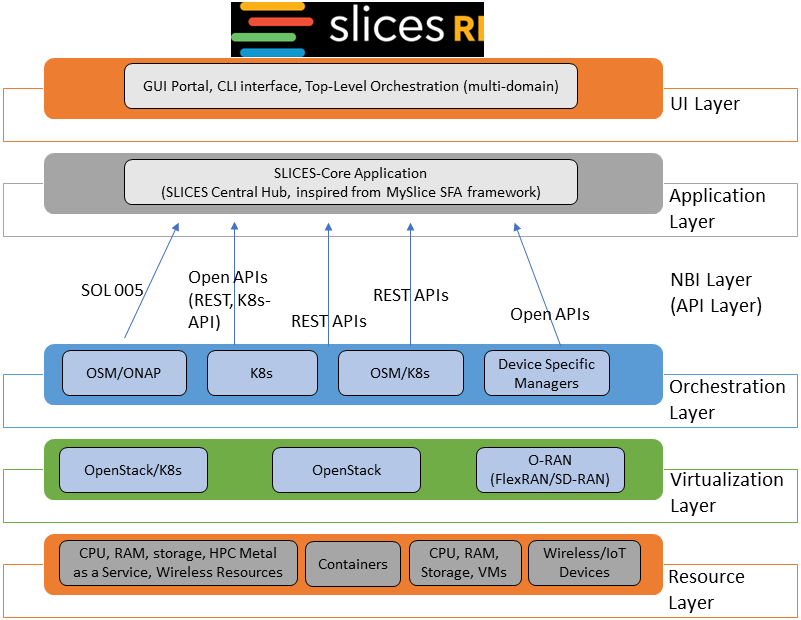}
	\caption{Layered architecture for SLICES}
	\label{fig:SLICES-Layered-architecture}
\end{figure*}

The SLICES architecture, illustrated in \Cref{fig:SLICES-Layered-architecture}, can be designed considering the limitations and challenges of existing federation-based architectures such as SFA. For example, the SLICES architecture could be designed by advancing the Slice-based Federation Architecture (SFA) and further enhancements are required to overcome the limitations and complexities to integrate wireless, edge and other experimental resources. We can consider a layer-based architecture as shown in \Cref{fig:SLICES-Layered-architecture}.

In this architecture, every component of SLICES testbed falls under a certain layer:
\begin{enumerate}
    \item \textit{Resource Layer}: It includes experimental resources such as CPU’s, RAM, storage, containers, VM’s, network, wireless, HPC and IoT devices;
    \item \textit{Virtualization Layer}: This layer includes cloud computing platforms (e.g., Openstack) that virtualize the underlying hardware resources and provide interfaces to the higher layers for programming/instantiating services over them. Examples of such programming interfaces are the ones defined by the O-RAN alliance (e.g., A1/E2 interfaces ), or the P4 programming abstractions for wired networks;
    \item \textit{Orchestration Layer}: It includes tools that orchestrate and instantiate services over the infrastructure equipment. Examples of such tools are OSM, ONAP and Kubernetes, mainly involved in NFV Management and Orchestration. It provides Network-Function-as-a-service and exposes northbound interfaces (NBI) APIs to be used by external entities; 
    \item \textit{NBI Layer}: This Layer defines the Open APIs that can be used by the SLICES application framework. Examples of such interfaces are the SOL005, the SOL004 from the ETSI NFV-MANO architecture \cite{etsi013} that can be found as the NBI interface of several MANO compliant tools, or even more generic ones, like TM-Forum  based APIs for service lifecycle control;
    \item \textit{Application Layer}: This Layer will host the SLICES-Core application, located at the SLICES central hub. It is responsible for managing all experimental resources that are exposed by lower layers, saved in the database and is further exposed to experimenters as a Service-Catalogue. It also exposes NBI API’s that can be used by a 3rd party orchestrator. The architecture of SLICES-Core application will start from components similar to MySlice V2 ~\cite{baron:hal-01804013}, and will be further enhanced at later stages;
    \item \textit{UI Layer}: This Layer defines the User Interface for the experimenters. It should abstract the experiments enough to make them more user friendly as possible. 
\end{enumerate}

The operation of the central-hub relies on the control of multiple-domains through the SLICES core application. Its operation resembles the functionality of a multi-domain orchestrator, that brings together different domains (in different locations, managed from different authorities) under the supervision of a single authority. The multi-domain orchestrator glues NFV, MEC and Cloud-Native orchestrators using API abstraction layers. Different groups of experimental resources on any of those testbeds might be virtualized and managed based on different technologies (e.g., VMs and containers). This in turn requires that multi-domain orchestrators use different orchestrators, managing different types of resources. For example, an NFVO and a MEC (Multi-access Edge computing Application Orchestrator-MEAO) are likely to be required in individual testbeds services, both managed concurrently through similar APIs from the same central entity.


The central SLICES core application shall include an abstraction API, used to trigger the required API invocation chains on different domain orchestrators when a high-level action is performed. A set of southbound clients is used in order to connect to NFV, MEC and Cloud-native local domain orchestrators. 

It is worth noticing that the different levels of access provided over the facility will also correspond to finer or coarser grain control over the deployment of experimental resources. More experienced users would be willing to control exactly which experimental resource to use from which facility at which site, while less-experienced users might not even know that their experiment is actually using heterogeneous resources composed out of different facilities spread across Europe. In the latter case, SLICES, through dedicated management components, will automatically assign resources to the users’ experiments, in order to optimize the overall utilization of the RI’s resources or simplify the work of the experimenter.


\section{The research Life Cycle}
\label{section:research}

Experimentally-driven research should be grounded on a solid methodology that is understood and implemented by other disciplines. This is somehow the ambition of the European EOSC initiative. As a consequence, SLICES does not target only the deployment of the instrument/facility but as importantly, addresses the full research life-cycle, including open data, data management and reproducibility.

Researchers and research stakeholders nowadays require that research data is made available for other researchers to examine, experiment and develop further. Additionally, preserving the data in conjunction with how conclusions from the data were drawn, accelerates the discovery process, enable easier reproducibility of the results and thus supports evidence. It is then necessary to develop policies and procedures for regulating the management and publication of research data in order to make them interoperable and widely available.

In Europe, it is recommended to conform with the European Open Science~\cite{euos} and Open Access policy \cite{euoa}, Open Research Data Pilot \cite{eu_data_pilot} and FAIR \cite{fair} principles in producing and managing research data. This requires defining appropriate metadata (including compatible experiment description) on the data produced by or integrated into the infrastructure with the objective to ensure eventually data accessibility, trustworthiness, reusability and interoperability with data produced by similar infrastructures/experiments for enabling complex experiments and multi-domain research. 
Alignments with the relevant recommendations such as the ones published by EOSC FAIRsFAIR \cite{fairsfair} project, GO FAIR initiative \cite{gofair} and RDA for FAIR data management \cite{rda2020fair}, and general European Open Access to research publications and Open Research Data Pilot policies, are of utmost importance.

The FAIR (Findable, Accessible, Interoperable, and Reusable) \cite{fair-principles} Data Principles were developed to be used as guidelines for data producers and publishers, with regards to data management and stewardship. One important aspect that differentiates FAIR from any other related initiatives is that they move beyond the traditional data and they place specific emphasis on automatic computation, thus considering both human-driven and machine-driven data activities. Since their publication, FAIR principles became widely accepted and used.
To this end, SLICES fully endorses and adopts the FAIR principles, acting as a catalyst to enable and foster the data-driven science and scientific data-sharing in this area.

Understanding the data collected and processed within SLICES becomes essential to understand data usage from the target user groups. This should allow to develop an appropriate information model that represents the data collected from the SLICES testbeds, experimental equipment and applications. We consider that the datasets generated by the usage of the SLICES hardware and software infrastructure can be roughly organized into five main categories:

\begin{itemize}
    \item[-] \textbf{Observational Data:}  collected using methods such as surveys (e.g. online questionnaires) or recording of measurements (e.g. through sensors). The data include mostly data related to signal or performance measurements, and network or service log data that allow for experiment evaluation and reproducibility. 
    \item[-] \textbf{Experimental Data: } where researchers introduce an intervention and study the effects of certain variables, trying to determine their impact.
    \item[-] \textbf{Simulation Data: } is generated by using computer models that simulate the operation of a real-world process or system. These may use observational data.
    \item[-] \textbf{Derived Data: } involves the analysis (e.g. cleaning, transformation, summarization, predictive modeling) of existing data, often coming from different datasets (e.g. the results of two experiments), to create a new dataset for a specific purpose. 
    \item[-] \textbf{Metadata:  } concerns data that provides descriptors about all categories of data mentioned above. This information is essential in making the discovery of data easier and ensuring their interoperability.
\end{itemize}

SLICES, as an open platform, promotes interoperability, thus non-proprietary, unencrypted, uncompressed, and commonly used by the research community formats should be adopted. In addition, SLICES end users should have the ability to decide on a suitable license and attach it to their data. 

Our preliminary estimations for SLICES include up to 5,000 users and their data, accounting for up to 50GB per user on the individual nodes and up to 1TB on the cloud. This provides us with a preliminary estimation of 0.25PB-1PB of data storage for all datacenters residing on SLICES nodes, and 5PB for the cloud-based datacenter.

As a consequence, SLICES will setup a data management framework to support the efficient and effective operation of the SLICES infrastructure. To accomplish this, the data management framework sets its own design goals, which are summarized below.

\begin{itemize}
    \item[-] \textbf{Data Governance: } A systemic and effective Data Governance structure to support the data management operations through a hierarchical structure with appropriate roles (e.g. Data Manager, Data Protection Officer and Metadata administrator), implement all related policies and processes, and adopt standards and leading practices.
    \item[-] \textbf{Data Architecture: } An agile Data Architecture that can perform efficiently to fulfill the SLICES infrastructure requirements, scales gracefully to accommodate for increased workloads, is flexible to integrate new processes and technologies, and is open to interact with other systems and infrastructures. 
    \item[-] \textbf{Data Quality:} Appropriate data transformation mechanisms to ensure Data Quality across multiple dimensions (e.g. accuracy, completeness, integrity), in order to improve data utility (e.g. further processing, analysis).
    \item[-] \textbf{Metadata:} Appropriate metadata management mechanisms to facilitate collaboration between users by providing the means to share their data and also support FAIR data.  
    \item[-] \textbf{Interoperability:  } Facilitate seamless interaction with other systems and infrastructures.
    \item[-] \textbf{Analytics:} Deployment of statistical, machine learning and artificial intelligence techniques to draw valuable insights from data and appropriate visualisation techniques to interpret them.
    \item[-] \textbf{Data Security:  } Mechanisms to protect data from unauthorized access and protect its integrity.
    \item[-] \textbf{Privacy: } Strict controls to manage the sharing of data, both internally and externally.
    
\end{itemize}


\section{Interoperability with EOSC and External Systems}
\label{section:Interoperability}

Since SLICES aims to provide a pan-European experimental research platform by jointly utilizing the geographically dispersed computing, storage and networking RIs, it is highly important that the different RIs interacting in the experimental workflow are interoperable with each other. Similarly, existing research needs to be accessible and directly pluggable to SLICES services and sites. For example, considering a MEC use case, compute, storage and networking resources from different RIs can be used. In such a scenario, it is necessary that resource description, availability, execution and data exchange are smooth. This can only be assured if a common interoperability framework is adopted across the SLICES ecosystem so that different subsystems have a common understanding of resources and data/metadata are on the same page with respect to the licensing, copyright and privacy requirements.
The SLICES infrastructure is designed to ensure compatibility and integration with EOSC and existing ESFRI infrastructures, and be ready to offer advanced ICT infrastructure services to other RIs and projects, with the special focus on the FAIR data management and exchange. 

\begin{figure*}[t]
    \centering
    \includegraphics{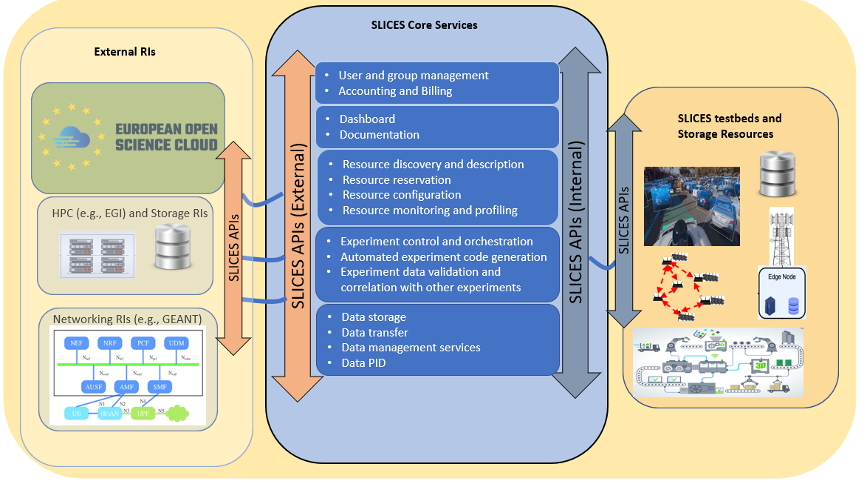}
	\caption{SLICES interconnection with European e-Infrastructures and digital infrastructures}
	\label{fig:SLICES interface to EOSC}
\end{figure*}


EOSC~\cite{eosc} has established itself as an important pillar in the implementation of Open science concept by accelerating the adoption of the FAIR data practices among researchers in the European Union. Integration of SLICES into European Research Infrastructure via EOSC will facilitate data sharing and reuse among SLICES partners and the larger European researchers’ community. Interoperability-focused integration of SLICES with EOSC will make it easier for SLICES users to reap the benefits of many services and tools pertaining to diverse scientific domains that are being developed around the EOSC ecosystem. 

Therefore, it is of utmost importance to design the integration framework of SLICES with EOSC in such a way that the data exchange between SLICES and EOSC is interoperable for scientific workflow management for data storage, processing and reuse. To this end, the recommendations of the EOSC interoperability framework  are considered in great detail for the design of the SLICES interoperability framework.

Interoperability is an essential feature of EOSC ecosystem as a federation of services and data exchange is unthinkable without interoperability among different EOSC constituents. The meaningful exchange and consumption of digital objects is necessary to generate value from EOSC, which can only be realized if different components of the EOSC ecosystem (software/machines and humans) have a common understanding of how to interpret and exchange them, what are the legal restrictions, and what processes are involved in their distribution, consumption and production. To facilitate this, EOSC interoperability framework (EOSC-IF)~\cite{eosc-if} is defined as a generic framework for all the entities involved in the development and deployment of EOSC.

To achieve this, a dedicated interface, coined SLICES-Interoperability Framework (SLICES-IF) shall be developed. The interface will be built upon the foundations led by the European Interoperability Reference Architecture (EIRA)~\cite{eira}, where interoperability is classified at four layers, namely: (i) technical, (ii) semantic, (iii) organizational; and (iv) legal. Although the target audience for EIRA (governance and administration) was very different from the SLICES stakeholders, core principles and objectives are similar. Additionally, the different components (in particular technical and semantic) of SLICES-IF would be chosen in such a way that SLICES is fully interoperable with EOSC for uninterrupted data exchange pertaining to use of EOSC services and research data by SLICES as well as to enable the publications of SLICES infrastructure, services and data through EOSC portal. More details about the SLICES-IF interface to EOSC and external RIs is provided in SLICES-Design Study Deliverable D4.2~\cite{slices-ds-d4.2}.

SLICES aims to allow its users (and interoperating platforms) to uniformly find, and access any object, such as data, services and software. To accomplish this, SLICES defines a hierarchical metadata structure, where each digital object is first described using a select set of common metadata attributes and then according to its type, the description is extended with a set of type-specific attributes.
The relevant information can be then accessed using SLICES authentication and authorization mechanisms. 



 \section{Full research life-cycle example}
 \label{section:Example}

\begin{figure*}[t]
    \centering
    \includegraphics[width=0.9\textwidth]{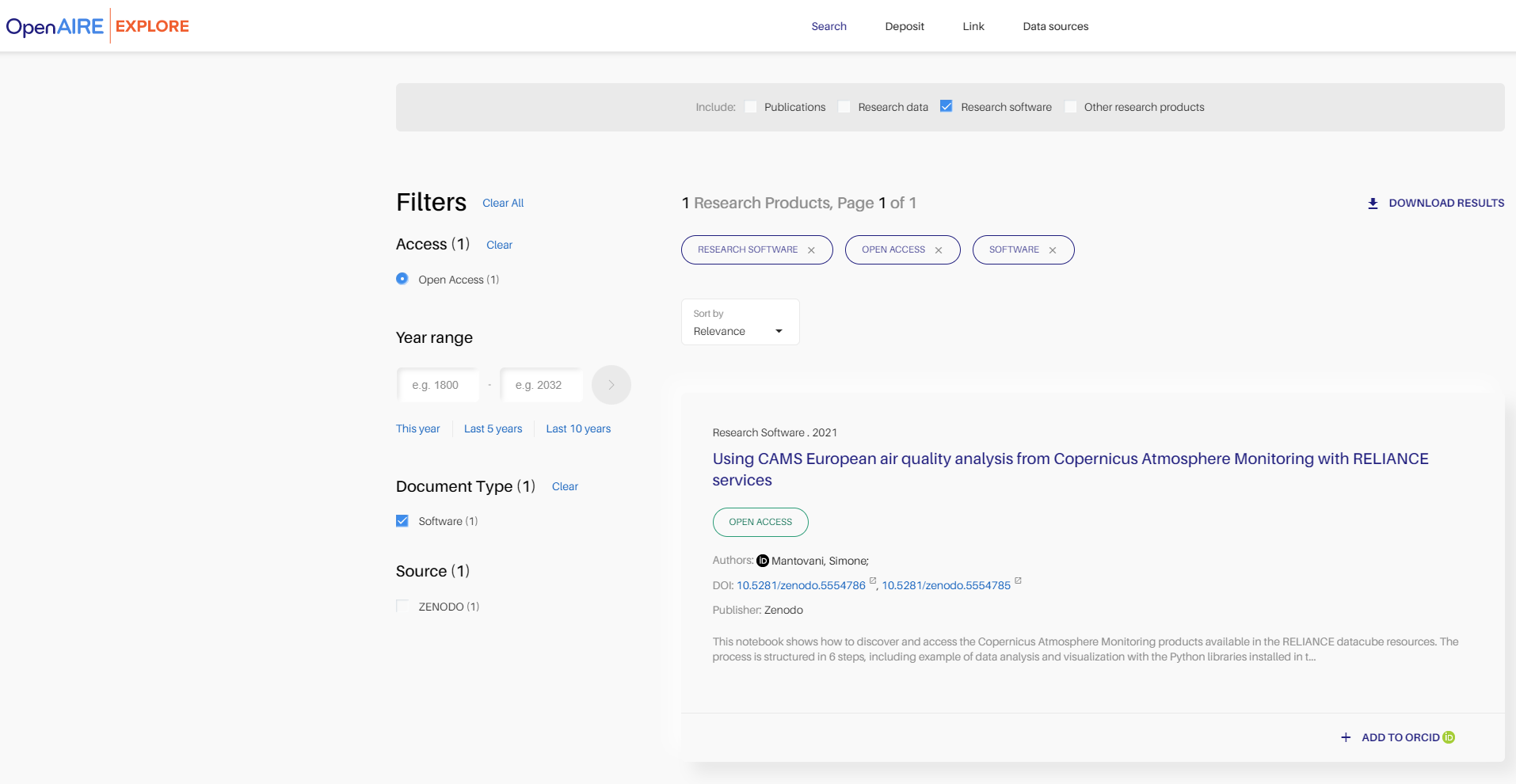}
	\caption{OpenAIRE Explore}
	\label{fig:OpenAIRE-EXPLORE}
\end{figure*}

\begin{figure*}[t]
    \centering
    \includegraphics[width=0.7\textwidth]{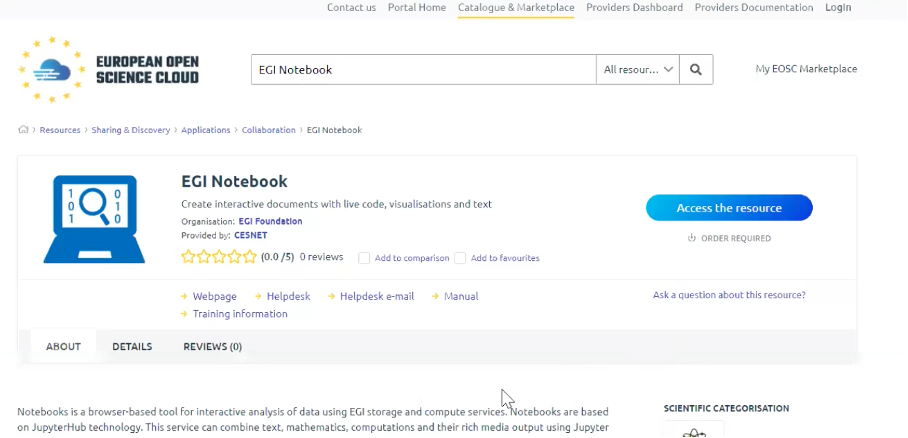}
	\caption{EGI Notebook}
	\label{fig:EGI-Notebook}
\end{figure*}

\begin{figure*}[t]
    \centering
    \includegraphics[width=\columnwidth]{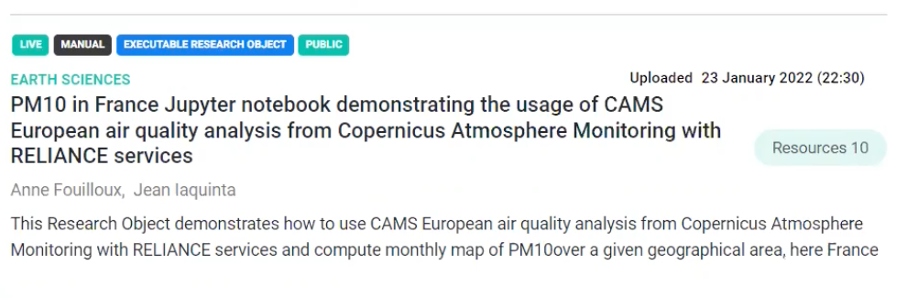}
	\caption{Jupyter notebook }
	\label{fig:Jupyter-notebook}
\end{figure*}

In order to realize the vision of FAIR research, supporting the full research lifecycle, lets consider a simple example borrowed from another field of research and illustrated by the Reliance project~\footnote{Reliance, https://www.reliance-project.eu/}. Reliance delivers a suite of innovative and interconnected services that extend EOSC’s capabilities to support the management of the research lifecycle within Earth Science Communities and Copernicus Users. Consider core services provided to the research community, that could be data, software publications, others. These core services are named after research objects that are for use by the experimenters and share by the experimenters. As an illustration, assume that you are doing some research related to the Copernicus air quality. You can go to the OpenAIRE~\footnote{OpenAIRE, https://explore.openaire.eu/} explorer and search for Copernicus quality. And you will find all the associated resources as described in \Cref{fig:OpenAIRE-EXPLORE}. 
You are looking for a software, because someone has produced a software taking research data as input and producing a map of the air quality in a given region as an output. You find the software and with the software comes a set of additional metadata. So for instance, it could be a Jupyter Notebook as in \Cref{fig:EGI-Notebook}.
You now have access to the software that will execute exactly what has produced this research data. What you are willing to do, at first, is to reproduce the results. On the other hand, you would like to take your own data, use the same process and produce your own new results. The last step is that you go to the service, which is named Rohub ~\footnote{Rohub, https://reliance.rohub.org/}. And then you bundle your different resources, like the Jupyter notebook that you have used, the data that you have exploited, and the output that you have produced. You can now publish this research outcome as your own contribution made available to the community, defined as PM 10 in \Cref{fig:Jupyter-notebook}. 
This full research-life cycle is really important, otherwise, the result that you produce cannot be published, because it simply cannot be reproduced. There exists data initiatives in our field, like the ACM Artifact Review level \cite{acm-artifact}, it is nice and ambitious. But it does not yet fully align with the best practices in other fields of research.

\section{Conclusion}
\label{section:conclusion}
It is a best practice in fundamental sciences to think about thought experiments that will validate the scientific assumptions. It is indeed a challenging endeavor to design a test platform to support networking and distributed research. Up to now, networking test-beds have tried to capture a variety of demands. However, very little has been done to cover the entire research and data lineage life-cycle. SLICES is the outcome of an effort to align the methodology to build such a platform in order to satisfy the key requirements of a scientific instrument. In Europe, ESFRI provides such a framework where most of the large research infrastructures are incubated, deployed and operated. 
The paper describes our continuous work aiming at designing the SLICES end-to-end reference architecture. It emphasized the analysis of the current demand from relevant ICT stakeholders, and the foundational principles on which it will be grounded. These principles, alongside with the current trends in resource management (resource programmability, network virtualization, resource disaggregation) have resulted in the wide adoption of several Management and Orchestration (MANO) frameworks for deploying experiments and applications over distributed infrastructures. The paper also discusses major open-source software that is used by the research community for networking at large, experiments that provides opportunities for SLICES. The integration and interoperability with EOSC infrastructure are also presented and the research life-cyle is illustrated. This work is based on the long experience of the participating members in managing and operating test platforms infrastructures to best serve our research community.

\section*{Acknowledgment}
The content of this paper is the outcome of a community work and we'd like to express our gratitude to all the SLICES partners and individuals who contributed to the discussion and preparation of the SLICES initiative (\url{https://slices-ri.eu}).
This project has received funding from the European Union's Horizon 2020 research and innovation program under Grant Agreement No 101008468 (SLICES-SC).

\bibliographystyle{IEEEtran}
\bibliography{header.bib, biblio.bib, sample-base.bib}

\begin{thebibliography}{10}
\providecommand{\url}[1]{#1}
\csname url@samestyle\endcsname
\providecommand{\newblock}{\relax}
\providecommand{\bibinfo}[2]{#2}
\providecommand{\BIBentrySTDinterwordspacing}{\spaceskip=0pt\relax}
\providecommand{\BIBentryALTinterwordstretchfactor}{4}
\providecommand{\BIBentryALTinterwordspacing}{\spaceskip=\fontdimen2\font plus
\BIBentryALTinterwordstretchfactor\fontdimen3\font minus
  \fontdimen4\font\relax}
\providecommand{\BIBforeignlanguage}[2]{{%
\expandafter\ifx\csname l@#1\endcsname\relax
\typeout{** WARNING: IEEEtran.bst: No hyphenation pattern has been}%
\typeout{** loaded for the language `#1'. Using the pattern for}%
\typeout{** the default language instead.}%
\else
\language=\csname l@#1\endcsname
\fi
#2}}
\providecommand{\BIBdecl}{\relax}
\BIBdecl

\bibitem{planetlab}
P.~Antoniadis \emph{et~al.}, ``{OneLab: An open federated facility for
  experimentally driven future internet research},'' in \emph{Proceedings of
  Conext 2010}.\hskip 1em plus 0.5em minus 0.4em\relax Springer, 2010, pp.
  1--12.

\bibitem{fed4fire}
T.~Wauters \emph{et~al.}, ``{Federation of internet experimentation facilities:
  architecture and implementation},'' in \emph{European Conference on Networks
  and Communications (EuCNC 2014)}, 2014.

\bibitem{federation}
\BIBentryALTinterwordspacing
P.~Antoniadis \emph{et~al.}, ``Federation of virtualized infrastructures:
  Sharing the value of diversity,'' in \emph{Proceedings of the 6th
  International COnference}, ser. Co-NEXT '10.\hskip 1em plus 0.5em minus
  0.4em\relax New York, NY, USA: Association for Computing Machinery, 2010.
  [Online]. Available: \url{https://doi.org/10.1145/1921168.1921184}
\BIBentrySTDinterwordspacing

\bibitem{orbit}
D.~Raychaudhuri \emph{et~al.}, ``{Overview of the ORBIT radio grid testbed for
  evaluation of next-generation wireless network protocols},'' in \emph{IEEE
  Wireless Communications and Networking Conference, 2005}, vol.~3.\hskip 1em
  plus 0.5em minus 0.4em\relax IEEE, 2005, pp. 1664--1669.

\bibitem{geni}
C.~Elliott, ``{GENI-global environment for network innovations.}'' in
  \emph{LCN}, 2008, p.~8.

\bibitem{eu-fire}
M.~Lemke, ``{The European FIRE Future Internet Research and Experimentation
  Initiative},'' in \emph{2009 5th International Conference on Testbeds and
  Research Infrastructures for the Development of Networks Communities and
  Workshops}, 2009, pp. 2--3.

\bibitem{sfa}
J.~Auge \emph{et~al.}, ``{Tools to foster a global federation of testbeds},''
  \emph{Computer Networks}, vol.~63, pp. 205--220, 2014.

\bibitem{pawr}
A.~Gosain, ``{Platforms for Advanced Wireless Research: Helping Define a New
  Edge Computing Paradigm},'' in \emph{Proceedings of the 2018 on Technologies
  for the Wireless Edge Workshop}, 2018, pp. 33--33.

\bibitem{5geve}
F.~Moggio \emph{et~al.}, ``{5G EVE a European platform for 5G Application
  deployment},'' in \emph{Proceedings of the 14th International Workshop on
  Wireless Network Testbeds, Experimental Evaluation \& Characterization},
  2020, pp. 124--125.

\bibitem{5genesis}
H.~Koumaras \emph{et~al.}, ``{5GENESIS: The Genesis of a flexible 5G
  Facility},'' in \emph{2018 IEEE 23rd International Workshop on Computer Aided
  Modeling and Design of Communication Links and Networks (CAMAD)}.\hskip 1em
  plus 0.5em minus 0.4em\relax IEEE, 2018, pp. 1--6.

\bibitem{5gvinni}
C.~Kalogiros \emph{et~al.}, ``{The potential of 5G experimentation-as-a-service
  paradigm for operators and vertical industries: The case of 5G-VINNI
  facility},'' in \emph{2019 IEEE 2nd 5G World Forum (5GWF)}.\hskip 1em plus
  0.5em minus 0.4em\relax IEEE, 2019, pp. 347--352.

\bibitem{onap}
``{Linux Foundation, ONAP – Open Network Automation Platform},'' [Online],
  \url{http://onap.org/}.

\bibitem{oran}
L.~Gavrilovska, V.~Rakovic, and D.~Denkovski, ``{From Cloud RAN to Open RAN},''
  \emph{Wireless Personal Communications}, pp. 1--17, 2020.

\bibitem{oai}
N.~Nikaein \emph{et~al.}, ``{OpenAirInterface: A flexible platform for 5G
  research},'' \emph{ACM SIGCOMM Computer Communication Review}, vol.~44,
  no.~5, pp. 33--38, 2014.

\bibitem{esfri}
``{European Strategy Forum on Research Infrastructures (ESFRI)},'' [Online],
  \url{https://www.esfri.eu/}.

\bibitem{sdn-nfv-review}
M.~S. Bonfim, K.~L. Dias, and S.~F. Fernandes, ``{Integrated NFV/SDN
  architectures: A systematic literature review},'' \emph{ACM Computing Surveys
  (CSUR)}, vol.~51, no.~6, pp. 1--39, 2019.

\bibitem{2018_cst_slicing}
I.~Afolabi \emph{et~al.}, ``Network slicing and softwarization: A survey on
  principles, enabling technologies, and solutions,'' \emph{IEEE Communications
  Surveys Tutorials}, vol.~20, no.~3, pp. 2429--2453, 2018.

\bibitem{2018_jsac_nfv}
X.~Cheng \emph{et~al.}, ``Network function virtualization in dynamic networks:
  A stochastic perspective,'' \emph{IEEE Journal on Selected Areas in
  Communications}, vol.~36, no.~10, pp. 2218--2232, 2018.

\bibitem{2022_6g_slicing}
W.~Wu \emph{et~al.}, ``Ai-native network slicing for 6g networks,'' \emph{IEEE
  Wireless Communications}, vol.~29, no.~1, pp. 96--103, 2022.

\bibitem{6953022}
F.~Rebecchi \emph{et~al.}, ``Data offloading techniques in cellular networks: A
  survey,'' \emph{IEEE Communications Surveys Tutorials}, vol.~17, no.~2, pp.
  580--603, 2015.

\bibitem{dressler22}
F.~Dressler \emph{et~al.}, ``V-edge: Virtual edge computing as an enabler for
  novel microservices and cooperative computing,'' \emph{IEEE Network}, 2022.

\bibitem{2018_access_slicing}
K.~Han \emph{et~al.}, ``Application-driven end-to-end slicing: When wireless
  network virtualization orchestrates with nfv-based mobile edge computing,''
  \emph{IEEE Access}, vol.~6, pp. 26\,567--26\,577, 2018.

\bibitem{2022_access_zero-touch}
H.~Chergui \emph{et~al.}, ``Toward zero-touch management and orchestration of
  massive deployment of network slices in 6g,'' \emph{IEEE Wireless
  Communications}, vol.~29, no.~1, pp. 86--93, 2022.

\bibitem{bonati2020open}
L.~Bonati \emph{et~al.}, ``Open, programmable, and virtualized 5g networks:
  State-of-the-art and the road ahead,'' \emph{Computer Networks}, vol. 182, p.
  107516, 2020.

\bibitem{o-ran-disaggregation}
M.~Polese \emph{et~al.}, ``{Understanding O-RAN: Architecture, Interfaces,
  Algorithms, Security, and Research Challenges},'' \emph{arXiv preprint
  arXiv:2202.01032}, 2022.

\bibitem{functional-splits}
L.~M. Larsen, A.~Checko, and H.~L. Christiansen, ``{A survey of the functional
  splits proposed for 5G mobile crosshaul networks},'' \emph{IEEE
  Communications Surveys \& Tutorials}, vol.~21, no.~1, pp. 146--172, 2018.

\bibitem{wypior2022open}
D.~Wypi{\'o}r, M.~Klinkowski, and I.~Michalski, ``Open ran—radio access
  network evolution, benefits and market trends,'' \emph{Applied Sciences},
  vol.~12, no.~1, p. 408, 2022.

\bibitem{cups}
J.~Kim, D.~Kim, and S.~Choi, ``{3GPP SA2 architecture and functions for 5G
  mobile communication system},'' \emph{ICT Express}, vol.~3, no.~1, pp. 1--8,
  2017.

\bibitem{ghosh20195g}
A.~Ghosh \emph{et~al.}, ``{5G evolution: A view on 5G cellular technology
  beyond 3GPP release 15},'' \emph{IEEE access}, vol.~7, pp.
  127\,639--127\,651, 2019.

\bibitem{rommer20195g}
S.~Rommer \emph{et~al.}, \emph{5G Core Networks: Powering
  Digitalization}.\hskip 1em plus 0.5em minus 0.4em\relax Academic Press, 2019.

\bibitem{5g-nr-3gpp}
3GPP, ``{3GPP TS 38.470 V17.0.0 (2022-04), 3rd Generation Partnership Project;
  Technical Specification Group Radio Access Network; NG-RAN; F1 general
  aspects and principles (Release 17)},'' 2022.

\bibitem{srs-ran}
``{srsRAN: a 4G/5G software radio suite},'' [Online],
  \url{https://github.com/srsran/srsRAN}.

\bibitem{5g-nrm}
3GPP, ``{3GPP TS 28.541 V17.6.0 (2022-03), 3rd Generation Partnership Project;
  Technical Specification Group Services and System Aspects; Management and
  orchestration; 5G Network Resource Model (NRM); Stage 2 and stage 3 (Release
  17)},'' 2022.

\bibitem{o-ran-controller}
L.~Bonati \emph{et~al.}, ``{Intelligence and learning in O-RAN for data-driven
  NextG cellular networks},'' \emph{IEEE Communications Magazine}, vol.~59,
  no.~10, pp. 21--27, 2021.

\bibitem{flexric}
R.~Schmidt, M.~Irazabal, and N.~Nikaein, ``{FlexRIC: an SDK for next-generation
  SD-RANs},'' in \emph{Proceedings of the 17th International Conference on
  emerging Networking EXperiments and Technologies}, 2021, pp. 411--425.

\bibitem{flexran}
X.~Foukas \emph{et~al.}, ``{FlexRAN: A flexible and programmable platform for
  software-defined radio access networks},'' in \emph{Proceedings of the 12th
  International on Conference on emerging Networking EXperiments and
  Technologies}, 2016, pp. 427--441.

\bibitem{onf-sd-ran}
ONF, ``{ONF Software Defined RAN (SD-RAN)},'' 2022, [Online],
  \url{https://opennetworking.org/sd-ran/}.

\bibitem{onf-aether}
------, ``{ONF AETHER - 5G Connected Edge platform},'' 2022, [Online],
  \url{https://opennetworking.org/aether/}.

\bibitem{berde2014onos}
P.~Berde \emph{et~al.}, ``{ONOS: towards an open, distributed SDN OS},'' in
  \emph{Proceedings of the third workshop on Hot topics in software defined
  networking}, 2014, pp. 1--6.

\bibitem{peterson2016central}
L.~Peterson \emph{et~al.}, ``{Central office re-architected as a data
  center},'' \emph{IEEE Communications Magazine}, vol.~54, no.~10, pp. 96--101,
  2016.

\bibitem{onf-omec}
ONF, ``{ONF OMEC - Open Mobile Evolved Core},'' 2022, [Online],
  \url{https://opennetworking.org/omec/}.

\bibitem{ts-23501}
3GPP, ``{3rd Generation Partnership Project; Technical Specification Group
  Services and System Aspects; System architecture for the 5G System (5GS);
  Stage 2 (Release 17)},'' 2022.

\bibitem{arouk20205g}
O.~Arouk and N.~Nikaein, ``{5G Cloud-Native: Network Management \&
  Automation},'' in \emph{NOMS 2020-2020 IEEE/IFIP Network Operations and
  Management Symposium}.\hskip 1em plus 0.5em minus 0.4em\relax IEEE, 2020, pp.
  1--2.

\bibitem{openstack}
T.~Rosado and J.~Bernardino, ``{An overview of OpenStack architecture},'' in
  \emph{Proceedings of the 18th International Database Engineering \&
  Applications Symposium}, 2014, pp. 366--367.

\bibitem{openvim}
R.~Mijumbi \emph{et~al.}, ``{Management and orchestration challenges in network
  functions virtualization},'' \emph{IEEE Communications Magazine}, vol.~54,
  no.~1, pp. 98--105, 2016.

\bibitem{k8s}
D.~Bernstein, ``{Containers and Cloud: From LXC to docker to Kubernetes},''
  \emph{IEEE Cloud Computing}, vol.~1, no.~3, pp. 81--84, 2014.

\bibitem{open5gs}
Open5GS, ``{Open Source project of 5GC and EPC (Release 16)},'' [Online],
  \url{https://open5gs.org/}.

\bibitem{oai-cn}
OAI, ``{OpenAirInterface Core Network},'' [Online],
  \url{https://gitlab.eurecom.fr/oai/cn5g}.

\bibitem{nextepc}
NextEPC, ``{NextEPC: Open Source EPC},'' [Online], \url{https://nextepc.org/}.

\bibitem{free5gc}
Free5GC, ``{Open Source implementation of 5G Core Network (3GPP Release 15 and
  beyond)},'' [Online], \url{https://www.free5gc.org/}.

\bibitem{magma}
Meta, ``{Magma: Communications Service Providers leverage Magma's open network
  core solution to connect people using LTE, 5G, Wi-Fi, and beyond.}''
  [Online], \url{https://www.facebook.com/connectivity/solutions/magma}.

\bibitem{ersue2013etsi}
M.~Ersue, ``{ETSI NFV management and orchestration-An overview},'' in
  \emph{Proc. of 88th IETF meeting}, 2013.

\bibitem{osm}
ETSI, ``{Open Source MANO},'' \emph{OSM home page -
  \url{https://osm.etsi.org/}}, 2016.

\bibitem{multi-tenant}
A.~{Mayoral} \emph{et~al.}, ``{Multi-tenant 5G Network Slicing Architecture
  with Dynamic Deployment of Virtualized Tenant Management and Orchestration
  (MANO) Instances},'' in \emph{ECOC 2016; 42nd European Conference on Optical
  Communication}, Sep. 2016, pp. 1--3.

\bibitem{openbaton}
G.~A. Carella and T.~Magedanz, ``{Open baton: a framework for virtual network
  function management and orchestration for emerging software-based 5G
  networks},'' \emph{Newsletter}, vol. 2016, p. 190, 2015.

\bibitem{docker}
C.~Anderson, ``{Docker [software engineering]},'' \emph{Ieee Software},
  vol.~32, no.~3, pp. 102--c3, 2015.

\bibitem{openshift}
S.~Picozzi, M.~Hepburn, and N.~O'Connor, \emph{{DevOps with Openshift: Cloud
  deployments made easy}}.\hskip 1em plus 0.5em minus 0.4em\relax " O'Reilly
  Media, Inc.", 2017.

\bibitem{mesos}
M.~Frampton, ``{Apache mesos},'' in \emph{Complete Guide to Open Source Big
  Data Stack}.\hskip 1em plus 0.5em minus 0.4em\relax Springer, 2018, pp.
  97--137.

\bibitem{macdavidSOSR2021}
R.~MacDavid \emph{et~al.}, ``A p4-based 5g user plane function,'' in \emph{SOSR
  '21: Proceedings of the ACM SIGCOMM Symposium on SDN Research (SOSR)}, 2021,
  pp. 162–--168.

\bibitem{kaltenberger2021}
F.~Kaltenberger, H.~Wang, and S.~Velumani, ``Performance evaluation of
  offloading ldpc decoding to an fpga in 5g baseband processing,'' in
  \emph{25th International ITG Workshop on Smart Antennas}, 2021, pp. 1--4.

\bibitem{oaixilinx2022}
\BIBentryALTinterwordspacing
 [Online]. Available:
  \url{https://www.xilinx.com/about/events/2022/mwc-2022.html}
\BIBentrySTDinterwordspacing

\bibitem{aerialSDK2022}
\BIBentryALTinterwordspacing
 [Online]. Available: \url{https://developer.nvidia.com/aerial-sdk}
\BIBentrySTDinterwordspacing

\bibitem{opennfv-edge}
``{OPEN NFV Edge project },'' [Online],
  \url{https://wiki.opnfv.org/display/EC/Edge+cloud}.

\bibitem{onap-project}
``{ONAP Project: Edge Automation through ONAP},'' [Online],
  \url{https://wiki.onap.org/display/DW/Edge+Automation+through+ONAP}.

\bibitem{openstack-project}
``{OpenStack Project: Edge Computing},'' [Online],
  \url{https://www.openstack.org/edge-computing/}.

\bibitem{lfedge-project}
``{LF EDGE Project: Building an Open Source Framework for the Edge},''
  [Online], \url{ https://www.lfedge.org /}.

\bibitem{5gcity-project}
``{5G City Project},'' [Online], \url{https://www.5gcity.eu/}.

\bibitem{openness-project}
``{Openness Project},'' [Online], \url{https://www.openness.org/}.

\bibitem{symec-project}
``{SYMEC Project},'' [Online], \url{https://www.symec.com.pl/}.

\bibitem{opnfv}
``{Linux Foundation Projects: OPNFV/Anuket},'' online],
  \url{[https://www.opnfv.org/}.

\bibitem{cncf}
``{Cloud Native Computing Foundation (CNCF)},'' online],
  \url{[https://www.cncf.io/}.

\bibitem{ocf}
``{Open Compute Project Foundation (OCP)},'' online],
  \url{[https://www.opencompute.org/}.

\bibitem{bernini2020multi}
G.~Bernini \emph{et~al.}, ``{Multi-domain orchestration of 5G vertical services
  and network slices},'' in \emph{2020 IEEE International Conference on
  Communications Workshops (ICC Workshops)}.\hskip 1em plus 0.5em minus
  0.4em\relax IEEE, 2020, pp. 1--6.

\bibitem{omf}
T.~Rakotoarivelo \emph{et~al.}, ``{OMF: a control and management framework for
  networking testbeds},'' \emph{ACM SIGOPS Operating Systems Review}, vol.~43,
  no.~4, pp. 54--59, 2010.

\bibitem{etsi013}
G.~ETSI, ``013: Network functions virtualisation (nfv); management and
  orchestration; os-ma-nfvo reference point--interface and information model
  specification.''

\bibitem{baron:hal-01804013}
\BIBentryALTinterwordspacing
L.~Baron \emph{et~al.}, ``{Next generation portal for federated testbeds
  MySlice v2: from prototype to production},'' {Fed4FIRE Engineering Conference
  - FEC2}, Oct. 2017, poster. [Online]. Available:
  \url{https://hal.archives-ouvertes.fr/hal-01804013}
\BIBentrySTDinterwordspacing

\bibitem{euos}
``{EU's open science policy},'' [Online],
  \url{https://ec.europa.eu/info/research-and-innovation/strategy/strategy-2020-2024/our-digital-future/open-science_en}.

\bibitem{euoa}
``{EU support for open access},'' [Online],
  \url{https://ec.europa.eu/info/research-and-innovation/strategy/strategy-2020-2024/our-digital-future/open-science/open-access}.

\bibitem{eu_data_pilot}
``{Open Research Data Pilot of the European Commission },'' 2017, [Online],
  \url{https://www.openaire.eu/what-is-the-open-research-data-pilot}.

\bibitem{fair}
B.~Mons \emph{et~al.}, ``{Cloudy, increasingly FAIR; revisiting the FAIR Data
  guiding principles for the European Open Science Cloud},'' \emph{Information
  Services \& Use}, vol.~37, no.~1, pp. 49--56, 2017.

\bibitem{fairsfair}
``{EOSC FAIRsFAIR project},'' [Online], \url{https://www.fairsfair.eu/}.

\bibitem{gofair}
``{EOSC GO FAIR initiative},'' [Online], \url{https://www.go-fair.org/}.

\bibitem{rda2020fair}
R.~F. D. M. M.~W. Group \emph{et~al.}, ``{FAIR Data Maturity Model:
  specification and guidelines},'' \emph{Research Data Alliance. DOI}, vol.~10,
  2020.

\bibitem{fair-principles}
M.~D. Wilkinson \emph{et~al.}, ``{The FAIR Guiding Principles for scientific
  data management and stewardship},'' \emph{Scientific data}, vol.~3, no.~1,
  pp. 1--9, 2016.

\bibitem{eosc}
P.~Ayris \emph{et~al.}, ``{Realising the European open science cloud},'' 2016.

\bibitem{eosc-if}
O.~Corcho \emph{et~al.}, \emph{{EOSC interoperability framework: Report from
  the EOSC Executive Board Working Groups FAIR and Architecture}}.\hskip 1em
  plus 0.5em minus 0.4em\relax European Commission, 2021.

\bibitem{eira}
M.~A. Wimmer, R.~Boneva, and D.~Di~Giacomo, ``{Interoperability governance: a
  definition and insights from case studies in Europe},'' in \emph{Proceedings
  of the 19th Annual International Conference on Digital Government Research:
  Governance in the Data Age}, 2018, pp. 1--11.

\bibitem{slices-ds-d4.2}
K.~Joshi \emph{et~al.}, ``{SLICES Design Study D4.2: SLICES infrastructure and
  services integration with EOSC and Open Science (initial proposal)},'' 2021,
  [Online],
  \url{http://slices-ds.eu/wp-content/uploads/2021/12/SLICES-DS\_D4.2.pdf}.

\bibitem{acm-artifact}
``{ACM Artifact Review and Badging},'' [Online],
  \url{https://www.acm.org/publications/policies/artifact-review-badging}.

\end{thebibliography}

\end{document}